\documentstyle[12pt,math,epsf,rotating,eepic]{article}
\begin{document}
\parskip=4mm
\parindent=0cm
\renewcommand\appendix{\par
  \setcounter{section}{0}%
  \setcounter{subsection}{0}%
  \setcounter{equation}{0}%
  \renewcommand\thesection{Appendix \Alph{section}.}
  \renewcommand{\theequation}{\Alph{section}.\arabic{equation}}}
\renewcommand{\theequation}{\arabic{section}.\arabic{equation}}
\renewcommand{\Re}{{\rm Re}\,}
\renewcommand{\Im}{{\rm Im}\,}
\newcommand{\pint}{\mbox{\scalebox{1.5}[1.5]
{\rotatebox[origin=c]{45}{-}}\hspace{-1.7ex}$\int$}}
\newcommand{\Pint}
{\mbox{\scalebox{2}[1]{\rotatebox[origin=c]{50}{--}}\hspace{-2.6ex}$\d\int$}}
\newcommand{\mod}{{\rm mod}\,}
\renewcommand{\arraystretch}{1.3}
\newcommand{\orintf}{\;\mbox{\setlength{\unitlength}{1mm}
\begin{picture}(4,2.4)(-1,-1)
\put(0,0){\makebox(0,0){$\displaystyle \int_{\Gamma'}$}}
\put(-0.8,0){\circle{3}}
\put(0.7,-0.8){\vector(0,1){1.6}}
\end{picture}}}
\newcommand{\orintg}[1]{\;\mbox{\setlength{\unitlength}{1mm}
\begin{picture}(4,2.4)(-1.5,-1)
\put(0,0){\makebox(0,0){$\displaystyle \int_{\partial #1}$}}
\put(-1.5,0){\circle{3}}
\put(0,-0.8){\vector(0,1){1.6}}
\end{picture}}}
\newcommand{\orinth}{\;\mbox{\setlength{\unitlength}{1mm}
\begin{picture}(3,2.4)(0,-1)
\put(0,0){\makebox(0,0){$\displaystyle \int_\Gamma$}}
\put(-0.4,0){\circle{3}}
\put(1.1,-0.8){\vector(0,1){1.6}}
\end{picture}}}
\today{}\hfill{UNIL-IPT-00-10}

\begin{center}
{\Large\bf The role of the input in Roy's equations for $\pi$-$\pi$ scattering}

G. Wanders

{\small Institut de Physique Th\'eorique, Universit\'e' de Lausanne\\
CH-1015 Lausanne, Switzerland\\
e-mail: gerard.wanders\@@ipt.unil.ch
}
\end{center}

\begin{abstract}The Roy equations determine the S- and P-wave 
$\pi$-$\pi$ phase shifts on a low energy interval. They allow the derivation 
of threshold parameters from experimental data. We examine the solutions of 
these equations that are in the neighborhood of a given solution by means of a 
linearization procedure. An updated survey of known results on the dimension 
of the manifold of solutions is presented. The solution is unique for a low 
energy interval with upper end at 800~MeV. We determine its response to small 
variations of the input: S-wave scattering lengths and absorptive parts above 
800~MeV. We confirm the existence of a universal curve of solutions in the 
plane of the S-wave scattering lengths and provide a control of the decrease 
of the influence of the input absorptive parts with increasing energy. A 
general result on the suppression of unphysical singularities at the upper end 
of the low energy interval is established and its practical implications are 
discussed.
\end{abstract}



\section{Introduction}
Low-energy $\pi$-$\pi$ scattering is a major testing ground of chiral 
perturbation theory~\cite{chpt}. Some of its coupling constants are directly 
related to the $\pi$-$\pi$ threshold parameters. At present this relation is 
established at the two-loop level~\cite{chpt,2loops}. As it is impossible to 
measure $\pi$-$\pi$ scattering at threshold, this relation cannot be exploited  
directly. A reliable extrapolation of the available experimental data down to 
threshold is required. Such an extrapolation is performed 
presently~\cite{anant} with the aid of the Roy equations~\cite{roy,mahoux}. 
These equations are based on the analyticity, the crossing symmetry and the 
unitarity of the $\pi$-$\pi$ partial wave amplitudes. The S- and P-wave Roy 
equations are solved in~\cite{anant} by means of elaborate numerical methods. 
In this work we discuss aspects of the problem which allow an analytical 
approach and our effort is complementary to the work in~\cite{anant}.

The Roy equations contain as input the S-wave scattering lengths, the S- and 
P-wave absorptive parts above an energy $E_0$ ( which will be called the 
``matching point") and driving terms coming from the higher partial waves. The 
Roy equations determine, at fixed elasticities, the S- and P-wave phase 
shifts below the matching point. This is a difficult, non-linear, problem that 
cannot be solved analytically. Here we restrict ourselves to questions that 
can be answered by linearization and which allow a partly analytic 
treatment. These concern the multiplicity of the solution and its sensitivity 
to small variations of the input.

Such questions have already been treated in~\cite{pomponiuw,epelew} in 
conjunction with the early phenomenological applications of the Roy 
equations~\cite{applicroy} and we can reduce the discussion of the multiplicity 
of the solution to the statement of our old results. The matching point used 
in~\cite{epelew} is at 1.13~GeV whereas the one used nowadays 
in~\cite{anant} is at $E_0=800$~MeV. The answers to our questions depend on 
the choice of $E_0$. The solution is non-unique if $E_0=1.13$~GeV and becomes 
unique when $E_0=800$~MeV. The response to variations of the input also 
depends strongly on the position of the matching point and our analysis 
in~\cite{epelew} has to be updated.

The Roy equations with arbitrarily chosen input make up a well defined 
mathematical problem. A peculiar feature of this problem is that its solutions 
exhibit unphysical singularities at the matching point [we exclude throughout 
matching points coinciding with an inelastic threshold]. The physical 
input\footnote{By physical input we denote the input corresponding to 
scattering amplitudes which would be measured in the absence of isospin 
violation. Ideally, our physical amplitudes are those provided by QCD; in 
practice they are given by the available analyses of experimental data 
assuming isospin symmetry.} is 
therefore a special one admitting at least one solution, the physical phase 
shifts, that is regular at $E_0$. Inputs with solutions regular at $E_0$ have 
been called ``analytic inputs" in~\cite{gasserw} in the context of simplified 
elastic one-channel Roy equations. The discussion of that class of inputs is 
extended here to the case of the complete coupled inelastic Roy equations. The 
main conclusion is that an analytic input admits a unique solution that is 
regular at the matching point. The non-uniqueness problem is thus evacuated.

Although non-uniqueness and singularities at $E_0$ are physically excluded, 
they show up in practical calculations because one is working with an 
approximate input which is not exactly an analytic one. An arbitrary variation 
of an analytic input produces a non-analytic one and induces singularities at 
$E_0$ even if the choice of $E_0$ guarantees uniqueness.

We find that one may stay close to an analytic input by correlating suitably 
the variations of two distinct pieces of the input. This comes mainly from the 
matching point at $E_0=800$~MeV that is near the $\rho$-meson mass. For 
instance, singularities at $E_0$ are largely suppressed by correlating 
variations of the isospin 0 and 2 S-wave scattering lengths $a_0^0$ and 
$a_0^2$. This confirms the existence of a physically acceptable family of 
solutions along a ``universal curve" in the 
$(a_0^0,a_0^2)$-plane~\cite{morgan1}. Similar suppressions of singularities take 
place if a localized variation of an input absorptive part is combined with a 
variation of one of the scattering lengths, $a_0^2$ for instance. The response 
to such variations provides information on the sensitivity of the phase shifts 
to the uncertainties on the input absorptive parts. We find a very weak 
sensitivity to the uncertainties above 1~GeV. All our results are in 
qualitative and quantitative agreement with those obtained numerically 
in~\cite{anant}.

The coupling between the S- and P-wave channels built into the Roy equations 
is a manifestation of crossing symmetry. The practical implications of this 
symmetry are not well understood and the effects of variations of the input 
might be expected to provide 
some insight. We find that this is not really the case. In our framework the 
response to a change of the input absorptive part in one channel is largest in 
the same channel but the responses in the other channels are not much 
smaller. All we may say is that crossing symmetry produces a substantial 
coupling of the three S- and P-wave channels, but we do not recognize very 
striking features.

The paper is organized as follows. The linearization procedure developed 
in~\cite{pomponiuw,epelew} is described, and the status of the uniqueness 
problem is outlined, in Section~2. Section~3 is devoted to the response to 
variations of the S-wave scattering lengths and the existence of a universal 
curve. The effects of correlated localized variations of input absorptive 
parts and variations of a scattering length are presented in Section~4. 
Variations of the driving terms are also briefly discussed in that section and 
our conclusions are displayed in Section~5. The fact that an analytic input 
admits only one solution that is regular at the matching point is a crucial 
result. We find it convenient to separate its proof from the presentation of 
phenomenological results and to explain it in Appendix~A. Our approximation 
scheme for the determination of linear responses is described in Appendix~B 
and Appendix~C gives a list of the kernels entering the S- and P-wave Roy 
equations.

\setcounter{equation}{0}
\section{Solution manifold of the S- and P-wave Roy equations} 

To set the stage we recall the main features of the S- and P-wave Roy 
equations~\cite{roy}. They relate the real and imaginary parts of the S- and 
P-wave $\pi$-$\pi$ scattering amplitudes at low energies, below the matching 
point $E_0$~:
\begin{equation}\label{2one}
\Re f_i(s)=(s-4)\sum_{j=0}^2{1\over \pi}\Pint_4^{s_0}{\rm d}x{1\over x-4}
\left[{\delta_{ij}\over x-s}+R_{ij}(s,x)\right]\Im f_j(x)+\phi_i(s),
\end{equation}
$i=0,1,2$. To lighten the writing, our notation differs from the standard 
one: $f_0$ and $f_2$ are the isospin $I=0$ and $I=2$ S-wave 
amplitudes and $f_1$ is the isospin $I=1$ P-wave. We return to the 
conventional notation $f_l^I$ in the presentation of final results. The 
variables $s$ and $x$ are 
squared total CM energies in units of $M_\pi^2$ ($M_\pi$ = pion mass, 
$s_0=(E_0/M_\pi)^2$). The equations~(\ref{2one}) contain singular diagonal 
Cauchy kernels and regular kernels $R_{ij}$ which are displayed in Appendix~C.

The $\phi_i$ are input functions
{\small
\begin{eqnarray}\label{2two}
\phi_i(s)&=&a_i+(s-4)\Bigl\{c_i(2a_0-5a_2)\\
&&\qquad +\sum_{j=0}^2{1\over \pi}\Pint_{s_0}^\infty
{\rm d}x{1\over x-4}\left[{\delta_{ij}\over x-
s}+R_{ij}(s,x)\right]A_j(x)+\psi_i(s)\Bigr\}.\nonumber
\end{eqnarray}}
In this equation $a_0$ and $a_2$ are the isospin 0 and 2 S-wave scattering lengths, 
$a_1=0$ here and 
\begin{equation}
c_0={1\over 12},\;c_1={1\over 72},\; c_2=-{1\over 24};
\end{equation}
the $A_i$ are the absorptive parts above the matching point~:
\begin{equation}
A_i(s)=\Im f_i(s),\qquad s\geq s_0,
\end{equation}
and the $\psi_i$ are so-called driving terms describing the contributions of 
the higher partial waves ($l\geq 2$). They have partial wave expansions 
converging in $[4,s_0]$ as long as $s_0<125.31$~\cite{mahoux}. The equations 
(\ref{2one}) constrain the S- and P-waves on $[4,s_0]$ at given 
input $(a_i,A_i,\psi_i)$. Unitarity implies
\begin{equation}\label{2five}
f_i(s)={1\over 2i\sigma(s)}\left(\eta_i(s)
{\rm e}^{2{\rm i}\delta_i(s)}-1\right),\quad \sigma(s)=\sqrt{1-{4\over s}},
\end{equation}
where $\delta_i$ is the channel $i$ phase shift and $\eta_i$ is the 
elasticity parameter ($0\leq \eta_i\leq 1$) which we incorporate into the 
input.

At given input the equations~(\ref{2one}) are coupled non-linear integral 
equations for the phase shifts $\delta_i$ on the interval $[4,s_0]$. To be 
acceptable, a solution of these equations has to provide absorptive parts 
below $s_0$ that join continuously the inputs $A_i$ at that point:
\begin{equation}\label{2six}
\lim_{s\nearrow s_0}{1\over 2\sigma(s)}
\left(1-\eta_i(s)\cos(2\delta_i(s))\right)=A_i(s_0).
\end{equation}
This boundary condition has to be added to the equations~(\ref{2one}).

The Roy equations being singular, the uniqueness of their solution is by no 
means guaranteed. We sum up the discussion of that point using the technique 
developed in~\cite{pomponiuw,epelew}. This technique will be our main tool 
throughout this article.

We assume we have a set of phase shifts $\delta_i$ satisfying the equations 
(\ref{2one}) and (\ref{2six}), the amplitudes $f_i$ being given by 
(\ref{2five}). We ask if these equations have other solutions $\delta'_i$ with 
the same input. If the $\delta'_i$ are infinitesimally close to the $\delta_i$ 
the differences $(\delta'_i-\delta_i)$ obey the linearized coupled equations
\begin{equation}\label{2seven}
\cos(2\delta_i(s))h_i(s)=\sum_j{1\over \pi}\Pint_4^{s_0}{\rm d}x{1\over x-4}
\left[{\delta_{ij}\over x-s}+R_{ij}(s,x)\right]\sin(2\delta_j(s))h_j(s),
\end{equation}
where
\begin{equation}\label{2eight}
h_i(s)={1\over \sigma(s)}\eta_i(s)(\delta'_i(s)-\delta_i(s)).
\end{equation}
The boundary conditions~(\ref{2six}) imply
\begin{equation}\label{2nine}
h_i(s_0)=0,
\end{equation}
i.e.~$\delta'_i(s_0)=\delta_i(s_0)$. The homogeneous equations~(\ref{2seven}) 
with boundary conditions~(\ref{2nine}) may have non-trivial solutions because 
of the presence of Cauchy kernels. The uniqueness or non-uniqueness of the 
$\delta_i$ depends on the existence of such solutions.

If the regular kernels $R_{ij}$ are omitted, the equations (\ref{2seven}) 
decouple and one recovers the one-channel problem discussed 
in~\cite{gasserw}. The existence of non-trivial solutions of this problem 
depends on the value of the phase shift $\delta_i$ at the matching point 
$s_0$. We assume that $\delta_i(s_0)>-\pi/2$. There is no solution if 
$-\pi/2<\delta_i(s_0)<\pi/2$. If $\delta_i(s_0)>\pi/2$, the general solution is
\begin{equation}\label{2ten}
h_i(s)=(s-4)G_i(s)P_i(s),
\end{equation}
where
\begin{equation}\label{2eleven}
G_i(s)=\left({s_0\over s_0-
s}\right)^{m_i}\exp\left[{2\over\pi}\Pint_4^{s_0}{\rm d}x
{\delta_i(x)\over x-s}\right]
\end{equation}
with
\begin{equation}\label{2twelve}
m_i=\left[{2\over\pi}\delta_i(s_0)\right].
\end{equation}
$[x]$ is the greatest integer smaller than $x$ (as in~\cite{gasserw}, $s_0$ is 
chosen in such a way that $\delta_i(s_0)$ is not an integral multiple of 
$\pi/2$). The last factor $P_i$ in the r.h.s. of (\ref{2ten}) is an arbitrary 
polynomial of degree $m_i-1$.

The general solution of the complete set of coupled equations (\ref{2seven}) 
has a form similar to (\ref{2ten}):
\begin{equation}\label{2thirt}
h_i(s)=(s-4)G_i(s)\left[P_i(s)+H_i(s)\right]
\end{equation}
with corrections $H_i$~\cite{epelew}.

The $P_i$ are again arbitrary poynomials of degree $m_i-1$: $m_i$ is given 
by~(\ref{2twelve}) if $\delta_i(s_0)>\pi/2$; $m_i=0$ and $P_i=0$ if 
$|\delta_i(s_0)|<\pi/2$. The functions $H_i$ are regular on $[4,s_0]$ and are 
solutions of a set of coupled non-singular integral equations:
\begin{eqnarray}
\lefteqn{\delta_{m_i,0}H_i(s)-{1\over \pi}\int_4^{s_0}{\rm 
d}x\,G_i(x)\sin(2\delta_i(x)){H_i(x)-H_i(s)\over x-s}}\quad && \nonumber\\
\hspace*{4mm}&=& \sum_j{1\over \pi}\int_4^{s_0}{\rm 
d}x\,R_{ij}(s,x)G_j(x)\sin(2\delta_j(x))[P_j(x)+H_j(x)].\label{2fourt}
\end{eqnarray} 

According to definition (\ref{2eleven}) we have
\begin{equation}\label{2fift}
G_i(s)\sim(s_0-s)^{\gamma_i}
\end{equation}
for $s\sim s_0$ with $\d \gamma_i={2\over \pi}\delta_i(s_0)-m_i$. This shows 
that $G_i$ vanishes at $s_0$ if $\delta_i(s_0)>0$ and diverges at that point 
if $\delta_i(s_0)<0$. Due to the regularity of $H_i$ at $s_0$ the boundary 
condition~(\ref{2nine}) is automatically fulfilled if $\delta_i(s_0)>0$. If 
$-\pi/2<\delta_i(s_0)<0$, $H_i$ has to vanish at $s_0$ (remember that $P_i=0$ 
in this case).

We now apply these results to the uniqueness problem of the physical $\pi-\pi$ 
S- and P-waves as solutions of the Roy equations (\ref{2one}). The input 
$(a_i,A_i,\psi_i,\eta_i)$ is identified with the physical one and we take the 
physical phase shifts as our master solution $\delta_i$ of the Roy equations. 
The physical isospin 0 S-wave and isospin 1 P-wave phase shifts being 
positive~\cite{morgan2}, we have $m_0\geq 0$, $m_1\geq 0$ and $\gamma_0>0$, 
$\gamma_1>0$. On the other hand, the isospin 2 S-wave phase shift $\delta_2$ 
is negative $(-\pi/2<\delta_2\leq 0$), $m_2=0$, $P_2=0$ and $\gamma_2<0$. 
Consequently, the three boundary conditions (\ref{2nine}) are satisfied if
\begin{equation}\label{2sixt}
H_2(s_0)=0.
\end{equation}

As solutions of the equations (\ref{2fourt}), the $H_i$ are linear functionals 
of the polynomials $P_0$ and $P_1$. Condition~(\ref{2sixt}) gives a 
homogeneous linear equation relating the coefficients of these polynomials and 
reduces by one the number of free parameters. If $m_0+m_1>1$ we are left with 
$m_0+m_1-1$ free parameters. There is no non-trivial solution if $m_0+m_1\leq 
1$. If $m_0+m_1>1$, the physical phase shifts are embedded in a 
$d$-dimensional manifold of solutions of the Roy equations with $d=m_0+m_1-1$. 
If $m_0+m_1\leq 1$, they form an isolated solution of these equations.

The actual values of $m_0$ and $m_1$ depend on the choice of the matching 
point $s_0$. Taking into account the known behaviour of the physical phase 
shifts~\cite{morgan2}, one finds four different situations when 
$E_0=\sqrt{s_0}\,M_\pi$ is lowered from 1.15~GeV to threshold.
\begin{enumerate}
\item \underline{1~GeV$< E_0<1.15$~GeV} 
In that interval, 
$\pi<\delta_0(s_0)<3\pi/2$, $\pi/2<\delta_1(s_0)<\pi$. This gives $m_0=2$, 
$m_1=1$ and $d=2$. The physical S- and P-waves are members of a two-parameter 
family of solutions of the Roy equations at fixed physical input and fixed 
phase shifts at $s_0$. The physical solution can be selected by imposing the 
physical values of the position and width of the $\rho$-meson.
\item \underline{860~MeV$< E_0<1$~GeV} 
We now have $\pi/2<\delta_i(s_0)<\pi$, $i=0,1$ and $m_0=m_1=1$, $d=1$. The 
polynomials $P_0$ and $P_1$ reduce to constants related by the equations 
(\ref{2sixt}). The physical amplitudes belong to a one-parameter family of 
solutions. The position of the $\rho$-meson can be used as a parameter.
\item \underline{780~MeV$< E_0<860$~MeV} 
In this interval $m_0=0$, $m_1=1$ and $d=0$ because $0<\delta_0(s_0)<\pi/2$, 
$\pi/2<\delta_1(s_0)<\pi$. The polynomial $P_0$ vanishes and $P_1$ is a 
constant which is set equal to zero by condition (\ref{2sixt}). The physical 
amplitudes form an isolated solution of the Roy equations. Position and shape 
of the $\rho$-resonance are determined by the input.
\item \underline{280~MeV$< E_0<780$~MeV}
Here $0<\delta_i(s_0)<\pi/2$, $i=0,1$, $m_0=m_1=0$ and both $P_0$ and $P_1$ 
vanish. The physical amplitudes again define an isolated solution.
\end{enumerate}

The above results concern the mathematical problem defined by equations 
(\ref{2one}), (\ref{2five}) and (\ref{2six}). Due to the behavior 
(\ref{2fift}) of the $G_i$ at $s_0$, the representation (\ref{2thirt}) implies 
that if there are solutions $\delta'_i$ in the neighborhood of $\delta_i$ they 
are singular at $s_0$ and exhibit cusps at that point. These singularities are 
unphysical because the choice of $s_0$ is arbitrary. In fact, all solutions of 
the Roy equations with arbitrary input are singular 
at $s_0$. The physical amplitudes being regular at $s_0$, the physical input 
has to be such that the corresponding Roy equations have at least one solution 
which is nonsingular at $s_0$ and coincides with the physical amplitudes. 
Among all possible inputs the physical input is a very special one: it is an 
an analytic input in the sense of Ref.~\cite{gasserw}. It has been shown 
there that in simplified one-channel Roy equations with analytic input there 
is only one solution which is regular at $s_0$. This crucial result is extended 
to the present realistic case in appendix~A.

We see that there is no non-uniqueness problem when working with the exact 
physical input. For instance in case~1 above, one could vary the position and 
width of the $\rho$-resonance in the two-parameter family of solutions. The 
singularities at $s_0$ would disappear at the physical values of these 
parameters. In practice, however, the physical input is only known 
approximately and one is not really working with an analytic input. Therefore 
singularities are present at $s_0$ and non-uniqueness cannot be avoided if 
$E_0>860$~MeV. We have to put up with these unpleasant features which are 
merely consequences of a deficient knowledge of the physical input.

From now on we choose the matching point used in the low-energy extrapolation 
based on the Roy equations performed in~\cite{anant}~: $E_0=800$~MeV 
($s_0=33$). Non-uniqueness is avoided but there are unwanted cusps at 
the matching point. It turns out that some of these cusps are in fact 
a helpful tool. Their suppression provides insights into the correlations 
constraining the scattering lengths of an analytic input. This will be 
illustrated repeatedly in this paper.

\setcounter{equation}{0}
\section{Varying the S-wave scattering lengths: universal curve}
We come now to our main topic, the linear response to small variations of the 
input. We proceed along the same lines as in the previous section. Starting 
from the solution $\delta_i$ with input $(a_i,A_i,\psi_i,\eta_i)$, we 
determine the solution $\delta'_i$ produced by a slightly modified input in 
linear approximation. To obtain quantitative results, we need a model for the 
$\delta_i$ which provides an acceptable representation of the physical phase 
shifts. We use the Schenk parametrization~\cite{schenk}:
\begin{equation}\label{3one}
\delta_i(s)=\tan^{-1}\left\{\sigma(s){4-z_i\over s-
z_i}\left[a_i+b_iq^2+c_iq^4\right]f_i(s)\right\},
\end{equation}
where
\begin{equation}\label{3two}
q^2={s\over 4}-1,\qquad f_i=\left\{\begin{array}{ll}
1 & \mbox{for }i=0,2,\\
q^2 & \mbox{for }i=1.
\end{array}\right.
\end{equation}
The values of the parameters are given in Table~1 and the phase shifts are 
shown in Fig.~1.

\begin{center}
\begin{tabular}{|c|c|c|c|c|}
\hline
&&&&\\[-6mm]
$i$ & $a_i$ & $b_i$ & $c_i$ & $z_i$\\
&&&&\\[-6mm]
\hline
0 & 0.200 & 0.245 & -0.0177 & 39.3\\
1 & 0.035 & 2.76$\cdot 10^{-4}$ & -6.9$\cdot 10^{-5}$ & 31.1\\
2 & -0.041 & -0.0730 & -3.2$\cdot 10^{-4}$ & -37.3\\
\hline
\end{tabular}
\end{center}
{\small Table 1: Values of the coefficients in the parametrization (\ref{3one}) of the 
physical S- and P-wave phase shifts.}
\vspace*{8mm}

The elasticities $\eta_i$ are very close to 1 below our matching point and 
will be set equal to 1 in all our numerical results.

In the present section, we vary only the S-wave scattering lengths:
\begin{equation}\label{3three}
a_i\to a'_i=a_i+\delta a_i,\qquad i=0,2.
\end{equation}
One of the main goals of the low-energy extrapolation of the experimental 
data lies in the determination of these scattering lengths. This cannot be 
achieved directly by 
solving the Roy equations because the scattering lengths enter into the input 
of these equations. However, they can be predicted in an indirect way because 
the physical input is an analytic one. Consequently, the scattering lengths 
are not independent of the other pieces of the input. If we know the physical 
$A_i$, $\psi_i$ and $\eta_i$ we may solve the Roy equations for arbitrary 
scattering lengths $a_i$. According to Proposition~1 in Appendix~A, their 
physical values are obtained by varying these $a_i$ until 
one arrives at a solution which is regular at $s_0$. In practice, when working 
with an approximation of the physical $A_i$, $\psi_i$ and 
$\eta_i$, the scattering lengths have to be varied until the corresponding 
solution of the Roy equations can be declared a good approximation of the 
solution of the problem with exact input. This is precisely the procedure used 
in~\cite{anant} and it is instructive to have an explicit control of the 
response to the variations~(\ref{3three}).

Our task is to determine functions $h_i$ defined as in 
equation~(\ref{2eight}). In the linearized scheme the 
representation~(\ref{2thirt}) is replaced by~\cite{epelew}
\begin{equation}\label{3four}
\hspace*{-5mm}h_i(s)=G_i(s)\left\{{\delta a_i\over G_i(4)}+(s-4)\left[(-
s)^{m_i}
c_i(2\delta a_0-5\delta a_2)+p\,\delta_{i,1}+H_i(s)\right]\right\}.
\end{equation}
\begin{center}
\begin{turn}{-90}
\epsfysize=8cm\epsfbox{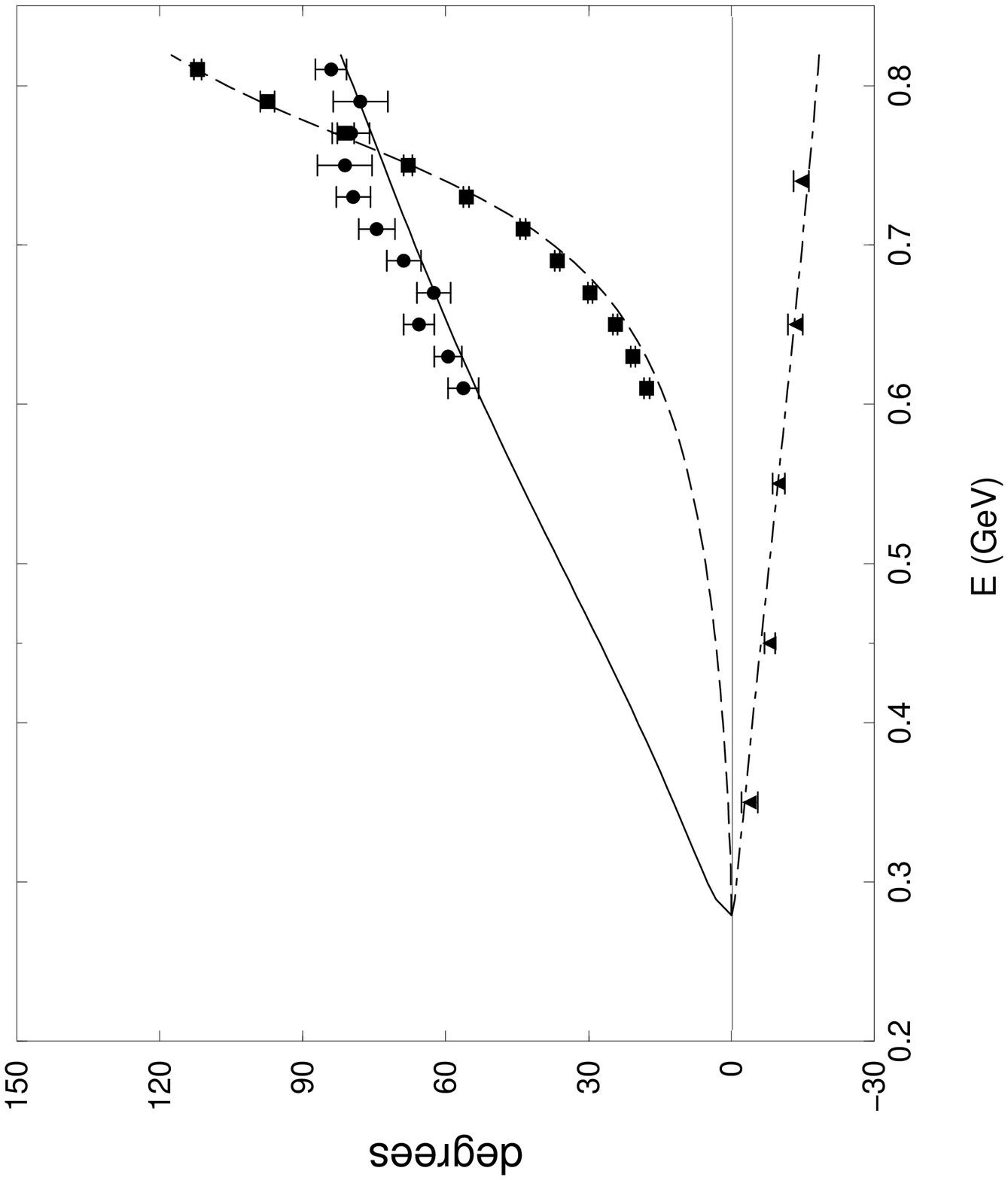}
\end{turn}
\end{center}
\setlength{\unitlength}{0.75mm}
{\small Figure 1: The $\pi$-$\pi$ S- and P-wave phase shifts according to the Ansatz 
(\ref{3one}) and data points obtained from analyses of experiments: 
$\delta_0^0$ and $\delta_1^1$ from~\cite{ochs} and $\delta_0^2$ 
from~\cite{hoogland}. 
\begin{picture}(14,2)(0,-2)\put(0,0){\line(1,0){14}}\end{picture} $\delta_0^0,$ 
\begin{picture}(14,2)(0,-2)\multiput(0,0)(3,0){5}{\line(1,0){2}}\end{picture} 
$\delta_1^1,$ 
\begin{picture}(14,2)(0,-2)\put(0,0){\line(1,0){2}}\put(3,0){\line(1,0){1}}
\put(5,0){\line(1,0){2}}\put(8,0){\line(1,0){1}}\put(10,0){\line(1,0){2}}
\put(13,0){\line(1,0){1}}
\end{picture} $\delta_0^2$.}

\vspace{4mm}
We recall that $m_0=m_2=0$, $m_1=1$, $P_0=P_2=0$ and $P_1$ is a constant which 
we call $p$. The functions $H_i$ are the solutions of inhomogeneous 
extensions of equations~(\ref{2fourt}). They have the form of 
equations~(\ref{Bone}) with
\begin{eqnarray}
Z_i &=& \sum_{j=0}^2Z_{ij},\nonumber\\
Z_{ij}(s)&=& {1\over\pi}\int_4^{s_0}{\rm 
d}x\,R_{ij}(s,x)G_j(x)\sin(2\delta_j(x))\times\nonumber\\
&&\quad \left\{{1\over (x-4)}{\delta a_j\over G_j(4)}+\left[(-x)^{m_j}
c_i(2\delta a_0-5\delta a_2)+p\,\delta_{j,1}\right]\right\}.\label{3six}
\end{eqnarray}

The solutions $H_i$ depend linearly on $p$ and this constant is fixed in such 
a way that $h_2(s_0)=0$, i.e.
\begin{equation}\label{3seven}
{\delta a_2\over G_2(4)}+(s_0-4)\left[c_2(2\delta a_0-5\delta a_2)
+H_2(s_0)\right]=0.
\end{equation}
We refer the reader to Ref.~\cite{epelew} for a derivation of equations 
(\ref{3four}) and (\ref{Bone}).

To obtain the $h_i$ we have to evaluate the molulating functions $G_i$ on the 
right-hand side of (\ref{3four}) and find the solution $H_i$ of (\ref{Bone}) 
and (\ref{3six}). The functions $G_i$ defined in (\ref{2eleven}) and obtained 
from the model (\ref{3one}) are shown in Fig.~2. The exponents $\gamma_i$ 
appearing in (\ref{2fift}) are 
\begin{equation}\label{3eight}
\gamma_0=0.89,\qquad \gamma_1=0.19,\qquad \gamma_2=-0.20.
\end{equation}
The small value of $\gamma_1$ comes from the fact that $s_0$ is close to 
$z_1$, the position of the $\rho$-resonance. This leads to the spectacular 
cusp of $G_1$ at $s_0$, seen in Fig.~2. As $h_2$ behaves as $(s_0-s)G_2$ at 
$s_0$, it is this product which is relevant and shown in Fig.~2. The exponents 
$\gamma_0$ and $\gamma_2+1$ are close to 1 and the cusps of $G_0$ and 
$(s_0-s)G_2$ are not visible in the figure.

\begin{center}
\begin{turn}{-90}
\epsfysize=8cm\epsfbox{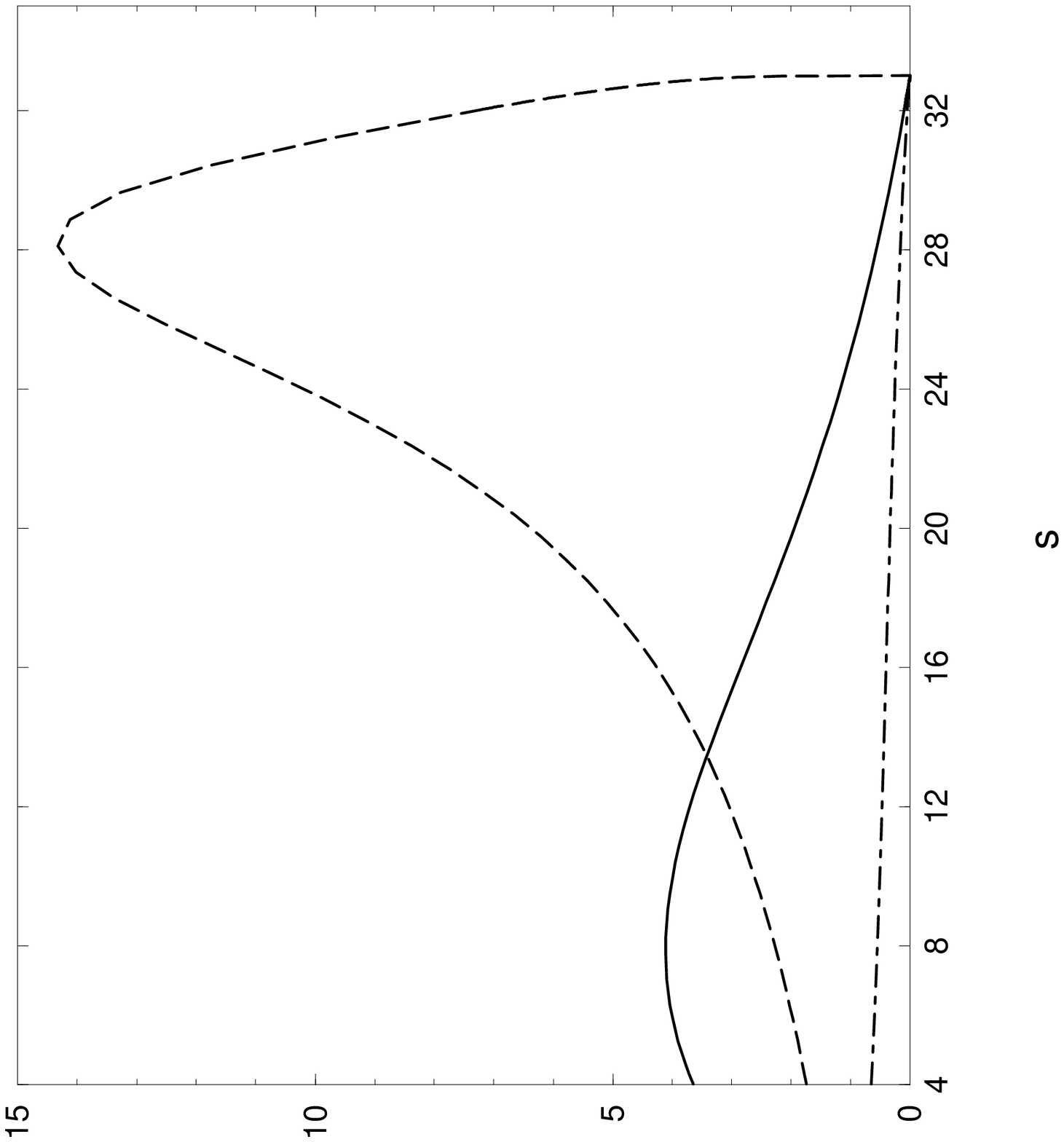}
\end{turn}
\end{center}
\nopagebreak
\setlength{\unitlength}{0.75mm}
{\small Figure 2: The functions $G_i$ defined in (\ref{2eleven}) appearing as factors in 
the responses to variations of the input. \begin{picture}(14,2)(0,-2)\put(0,0)
{\line(1,0){14}}\end{picture} $G_0,$ 
\begin{picture}(14,2)(0,-2)\multiput(0,0)(3,0){5}{\line(1,0){2}}\end{picture} 
$G_1,$ 
\begin{picture}(14,2)(0,-2)\put(0,0){\line(1,0){2}}\put(3,0){\line(1,0){1}}
\put(5,0){\line(1,0){2}}\put(8,0){\line(1,0){1}}\put(10,0){\line(1,0){2}}
\put(13,0){\line(1,0){1}}
\end{picture} $\d {s_0-s\over s_0}G_2$.}

\vspace{4mm}
The $H_i$ are slowly varying and (\ref{3four}) tells us that $h_1$ has a sharp 
cusp at $s_0$. The $\delta_i$ defined in (\ref{3one}) are models of the 
physical phase shifts: they are regular at $s_0$ and meant to be produced by 
an analytic input $(a_i,A_i,\psi_i,\eta_i)$. We see that $\delta'_1$ 
obtained from (\ref{2eight}) has a sharp cusp at $s_0$. This singular behavior 
is a visible signal that the modified input 
$(a_i+\delta a_i,A_i,\psi_i,\eta_i)$ is no longer an analytic one. The 
differences $(\delta'_i-\delta_i)$ are linear in $\delta a_0$ and $\delta 
a_2$:
\begin{equation}\label{3nine}
\delta'_i(s)-\delta_i(s)=G_i(s)\left[f_{i0}(s)\delta a_0+f_{i2}(s)\delta 
a_2\right],
\end{equation}
where $f_{i0}$ and $f_{i2}$ are regular at $s_0$. We see that the cusp of 
$\delta'_1$ is suppressed if
\begin{equation}\label{3ten}
f_{10}(s_0)\delta a_0+f_{12}(s_0)\delta a_2=0.
\end{equation}
There is a direction in the $(a_0,a_2)$-plane along which $\delta'_1$ has no 
cusp. There is still a singularity at $s_0$ but no infinite slope. This 
indicates that the input $(a_i+\delta a_i,A_i,\psi_i,\eta_i)$ is close to an 
analytic input if condition (\ref{3ten}) is satisfied. An analytic input is not 
an isolated one: it is transformed into new analytic inputs by suitably 
correlated variations of its ingredients. Our finding and the fact that 
$\delta'_0$ and $\delta'_2$ have no visible cusps show that the physical input 
is transformed into nearly analytic inputs by variations of the scattering 
lengths obeying (\ref{3ten}). One can show that variations of the scattering 
lengths alone cannot transform an analytic input exactly into an analytic one. 
A movement along a direction in the $(a_0,a_2)$-plane has to be accompanied by 
modifications of the remaining pieces of the input if one wants to keep it 
exactly analytic. Our results show that these modifications are small and we 
confirm at the local level the existence of a one-parameter family of nearly 
analytic inputs along a universal curve, $a_2=a_2(a_0)$ in the 
$(a_0,a_2)$-plane~\cite{morgan1}.

To go beyond qualitative results, equations~(\ref{Bone}) have to be solved 
and we apply the approximation scheme described in Appendix~B. In this scheme, 
the $H_i$ have the form
\begin{equation}\label{3eleven}
H_i(s)\simeq\left({s\over s_0}\right)^{m_i}\left(\hat{H}_{i,0}(s)\delta 
a_0+\hat{H}_{i,2}(s)\delta a_2+\hat{H}_{i,3}(s)p\right),
\end{equation}
where the $\hat{H}_{i,k}$ are second-degree polynomials. Once we have 
determined $\hat{H}_{i,k}$ and when $p$ has been fixed by (\ref{3seven}), we 
find that the ratio $\delta a_2/\delta a_0$ defined in (\ref{3ten}), for which 
the cusp in $\delta'_1$ is suppressed, is equal to 0.197. In fact the ratio 
$f_{1,0}(s)/f_{1,2}(s)$ is nearly constant and equal to its value at $s_0$ on 
the whole interval $[4,s_0]$. We identify the ratio 0.197 with the slope of 
the universal curve at $(a_0,a_2)$: it coincides with the slope found 
in~\cite{anant}. There is a strong compensation of the two terms in the 
right-hand side of equation~(\ref{3nine}) when $i=1$ if one moves along the 
universal curve. This compensation is maximally removed in the orthogonal 
direction where $\delta a_2/\delta a_0=-5.08$. This is illustrated in Fig.~3 
which displays the relative phase-shift differences $(\delta'_i-
\delta_i)_\Vert/\delta_i$ and $(\delta'_i-\delta_i)_\perp/\delta_i$. The 
differences $(\delta'_i-\delta_i)_\Vert$ are obtained when the point 
$(a_0,a_2)$ moves along the universal curve
\begin{equation}\label{3twelve}
\delta a_0=\delta a_\Vert\,\cos\theta_\Vert,\qquad\delta a_2=\delta 
a_\Vert\,\sin\theta_\Vert,
\end{equation}
with $\theta_\Vert=\tan^{-1}0.197=11^\circ$. The differences 
$(\delta'_i-\delta_i)_\perp$ are obtained in response to a displacement 
$(\delta a_0,\delta a_2)$ normal to the universal curve, $\delta a_0$ and 
$\delta a_2$ being given by (\ref{3twelve}) with $\delta a_\Vert$ replaced by 
$\delta a_\perp$ and $\theta_\Vert$ replaced by $\theta_\perp=101^\circ$.

\begin{minipage}{9cm}
\begin{turn}{-90}
\epsfxsize=8cm\epsfbox{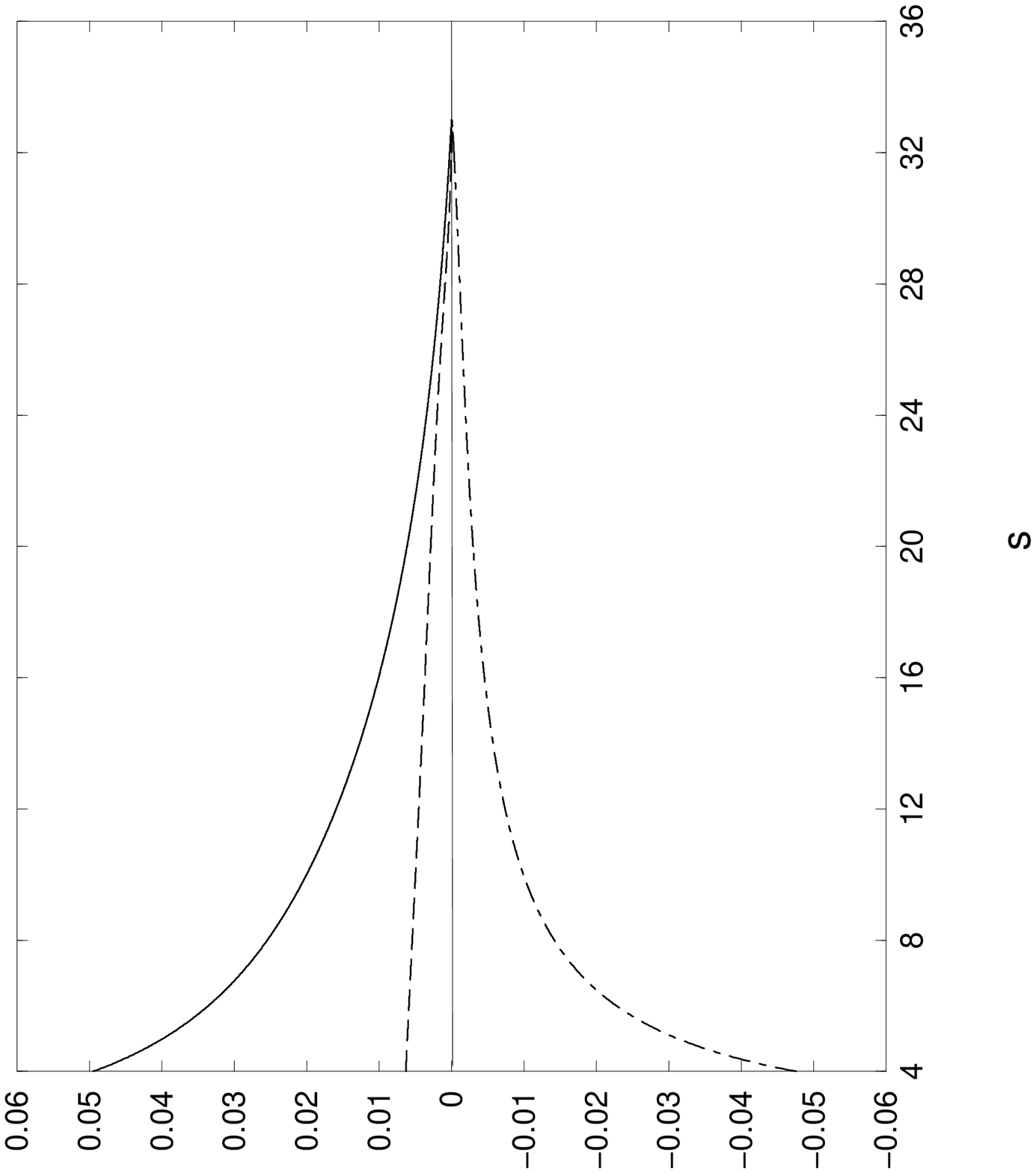}
\end{turn}
\end{minipage}
\setlength{\unitlength}{0.75mm}
\hfill\parbox{3.6cm}{$\mbox{ }$\\[8mm]
\footnotesize\begin{picture}(14,2)(0,-2)\put(0,0){\line(1,0){14}}\end{picture} 
$(\delta_0^{0'}-\delta_0^0)_\Vert/\delta_0^0,$\\[8mm] 
\begin{picture}(14,2)(0,-2)\multiput(0,0)(3,0){5}{\line(1,0){2}}\end{picture} 
$(\delta_1^{1'}-\delta_1^1)_\Vert/\delta_1^1,$\\[8mm]  
\begin{picture}(14,2)(0,-2)\put(0,0){\line(1,0){2}}\put(3,0){\line(1,0){1}}
\put(5,0){\line(1,0){2}}\put(8,0){\line(1,0){1}}\put(10,0){\line(1,0){2}}
\put(13,0){\line(1,0){1}}
\end{picture} $(\delta_0^{2'}-\delta_0^2)_\Vert/\delta_0^2$.\\[18mm]
Figure 3a}

\vspace{8mm}
\begin{minipage}{9cm}\begin{turn}{-90}
\epsfxsize=8cm\epsfbox{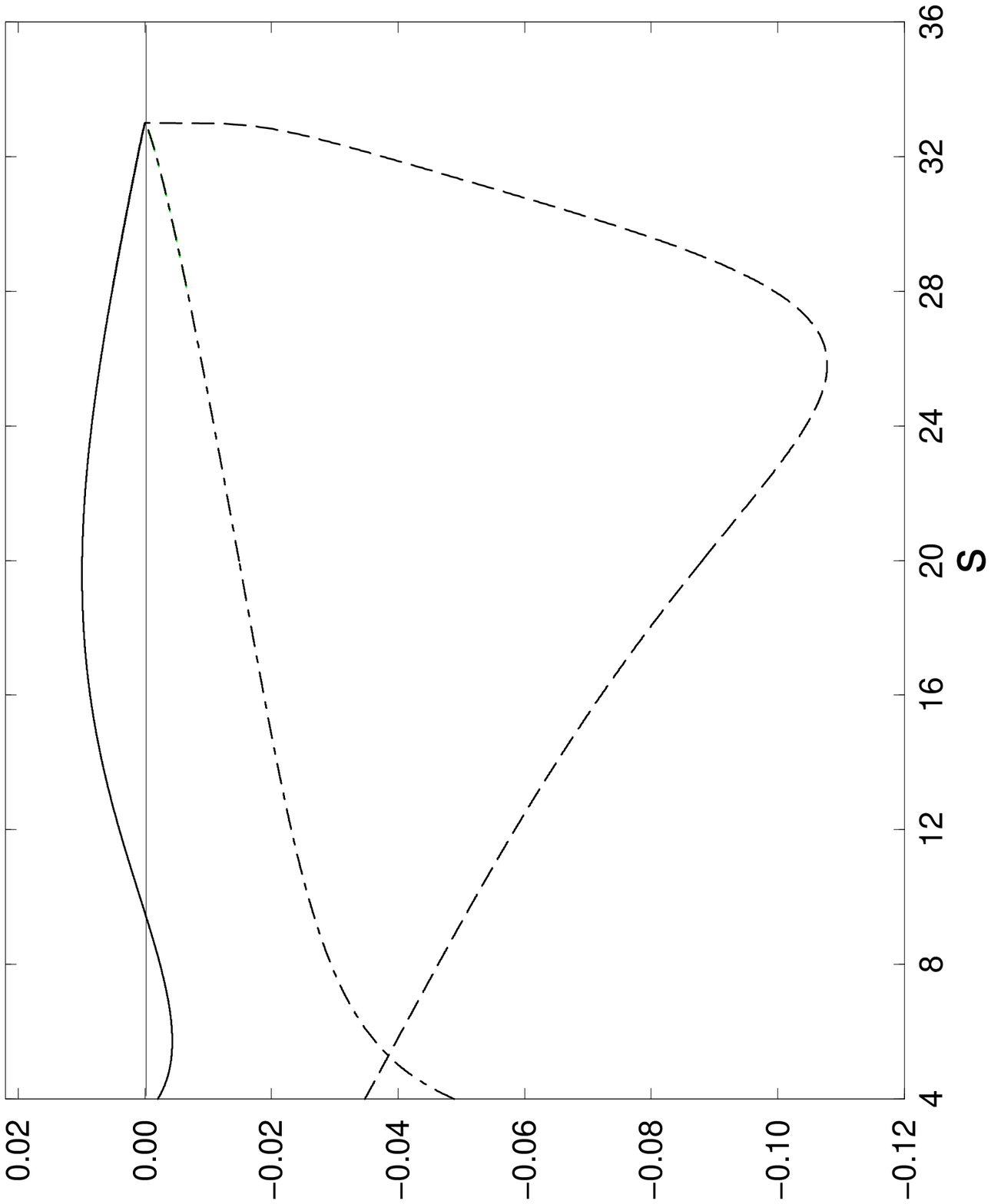}
\end{turn}\end{minipage}
\hfill\parbox{3.6cm}{$\mbox{ }$\\[8mm]
\footnotesize\begin{picture}(14,2)(0,-2)\put(0,0){\line(1,0){14}}\end{picture} 
$(\delta_0^{0'}-\delta_0^0)_\perp/\delta_0^0,$\\[8mm] 
\begin{picture}(14,2)(0,-2)\multiput(0,0)(3,0){5}{\line(1,0){2}}\end{picture} 
$(\delta_1^{1'}-\delta_1^1)_\perp/\delta_1^1,$\\[8mm]  
\begin{picture}(14,2)(0,-2)\put(0,0){\line(1,0){2}}\put(3,0){\line(1,0){1}}
\put(5,0){\line(1,0){2}}\put(8,0){\line(1,0){1}}\put(10,0){\line(1,0){2}}
\put(13,0){\line(1,0){1}}
\end{picture} $(\delta_0^{2'}-\delta_0^2)_\perp/\delta_0^2$.\\[18mm]
Figure 3b}

\nopagebreak
{\small Figure 3: Relative responses $(\delta_l^{I\,'}-\delta_l^I)/\delta_l^I$ to 
variations of the S-wave scattering lengths. 
(a) displacement (\ref{3twelve}) along the universal curve, 
$\delta a_\Vert=0.05 a_0^2$.\linebreak
(b) displacement orthogonal to the universal curve, $\delta a_\perp=0.05 
|a_0^2|$.}

\vspace{4mm}
To assess the quality of the results displayed in Fig.~3, Table~2 gives the 
values of the $\chi_i$ defined in (\ref{Bten}) in the parallel and orthogonal 
directions. The values of the $\chi_i^\perp$ are acceptable whereas those for 
$\chi_i^\Vert$ are surprisingly small.

\begin{center}
\begin{tabular}{|c|c|c|c|}
\hline
&&&\\[-6mm]
$i$ & 0 & 1 & 2\\
&&&\\[-6mm]
\hline
$\chi_i^\Vert$ & $3.7\cdot 10^{-4}$ & $1.9\cdot 10^{-3}$ & $2.9\cdot 10^{-4}$\\
$\chi_i^\perp$ & $9.5\cdot 10^{-3}$ & $3.1\cdot 10^{-2}$ & $1.8\cdot 10^{-2}$\\
\hline
\end{tabular}
\end{center}
{\small Table 2: Accuracy of the approximate values of $(\delta'_i-\delta_i)_\Vert$ 
and $(\delta'_i-\delta_i)_\perp$. The mean relative quadratic discrepancies 
$\chi_i$ are defined in (\ref{Bten}).}
\vspace*{8mm}

The relative variations of the phase shifts in the direction of the universal 
curve are decreasing functions of $s$. The S-waves have peaks at threshold, 
whose sizes are dictated by the values of $\delta a_0$ and$\delta a_2$. The 
pattern in the orthogonal direction is different and more complicated. The 
effects of the variation of the scattering length spread over the whole 
interval $[4,s_0]$ and $(\delta'_1-\delta_1)_\perp/\delta_1$ has a cusp which 
cannot be overlooked.

The overall size of the variations $(\delta'_i-\delta_i)_\perp$ is 
significantly larger than that of the corresponding 
$(\delta'_i-\delta_i)_\Vert$. To characterize this fact quantitatively we 
evaluate the mean values of the absolute ratios over the interval $[4,s_0-2]$ 
($s_0-2$ instead of $s_0$ as upper limit to avoid effects of the cusp in 
$(\delta'_1-\delta_1)_\perp$)
\begin{equation}\label{3thirt}
\rho_i=\left\langle\left|{(\delta'_i-\delta_i)_\perp\over
(\delta'_i-\delta_i)_\Vert}\right|\right\rangle.
\end{equation}
We take $\delta a_\perp=\delta a_\Vert$, i.e.~we assume that the same distance  
is covered along and perpendicularly to the universal curve, and find
\begin{equation}\label{3fourt}
\rho_0=15.7,\qquad \rho_1=217,\qquad \rho_2=18.6.
\end{equation}
These large values reflect the sharp definition of the universal curve 
obtained in~\cite{anant}.

\setcounter{equation}{0}
\section{Combined variations of input absorptive parts, driving terms and 
scattering length}
This section is mainly devoted to variations $\delta A_i$ of the absorptive 
parts $A_i$ above the matching point. As shown in~\cite{epelew} the response is 
obtained from the following Ansatz for the functions $h_i$ in (\ref{2eight}):
\begin{equation}\label{4one}
h_i(s)=G_i(s)(s-4)[p\,\delta_{i,1}+F_i(s)+H_i(s)],
\end{equation}
where
\begin{equation}\label{4two}
F_i(s)={s^{m_i}\over \pi}\int_{s_0}^\infty{\rm d}x\,{1\over x^{m_i}}
{1\over x-4}{1\over x-s}{\delta A_i(x)\over G_i(x)}.
\end{equation}
The functions $H_i$ on the right-hand side of eq.~(\ref{4one}) are solutions 
of the equations (\ref{Bone}) with
\begin{equation}\label{4three}
Z_i(s)=\sum_{j=0}^2Z_{ij}(s),
\end{equation}
where
\begin{equation}\label{4four}
Z_{ij}(s)=Y_{ij}(s)+{1\over\pi}\int_4^{s_0}{\rm 
d}x\,R_{ij}(s,x)\,\sin(2\delta_j(x))G_j(x)\left[p\,\delta_{j,1}+F_j(x)\right]
\end{equation}
and
\begin{equation}\label{4five}
Y_{ij}(s)={1\over \pi}\int_{s_0}^\infty{\rm d}x\,{1\over x-4}R_{ij}(s,x)\delta 
A_j(x).
\end{equation}
Equations (\ref{4two})-(\ref{4five}) have been derived in~\cite{epelew}. For 
simplicity we assume that the $\delta A_i$ vanish at $s_0$ and the boundary 
condition (\ref{2nine}) remains unchanged: $h_i(s_0)=0$.

Using the analyticity properties of the kernels $R_{ij}$ referred to in 
Appendix~B, the integral in the right-hand side of eq.~(\ref{4four}) can be 
transformed to give
\begin{equation}\label{4six}
Z_{ij}(s)=p\,Q_i(s)\delta_{j,1}+\int_{s_0}^\infty {\rm d}u\,M_{ij}(s,u){1\over 
u-4 }{\delta A_j(u)\over G_j(u)},
\end{equation}
where $Q_i$ is a known polynomial and
\begin{equation}\label{4seven}
M_{ij}(s,u)=-{1\over 2{\rm i}\pi}\orintf {\rm 
d}x\,R_{ij}(s,x)\bar{G}_j(x)\left({x\over u}\right)^{m_i}{1\over x-u}.
\end{equation}
The contour $\Gamma'$ encircles the segment $[-(s-4),0]$. Formula 
(\ref{4seven}) affords an explicit evaluation of $M_{ij}$ once the $\bar{G}_j$ 
have been approximated by polynomials, as in appendix~B. The functions $H_i$ 
are determined by applying the method of that Appendix. 
The condition $h_2(s_0)=0$ fixes $p$ as a linear functional of the $\delta 
A_i$ and the differences $\delta'_i-\delta_i$ resulting from (\ref{2eight}) 
and (\ref{4one}) can be written as
\begin{equation}\label{4eight}
\delta'_i(s)-\delta_i(s)=\sum_{j=0}^2\int_{s_0}^\infty{\rm 
d}u\,K_{ij}(s,u)\delta A_j(u).
\end{equation}
The kernel $K_{ij}(u)$ gives the effect on the channel $i$ phase shift of a 
variation of the channel $j$ absorptive part at point $u$ ($u>s_0$). It 
follows from (\ref{4one}) that $K_{ij}(s,u)$ is proportional to $G_i(s)$ and 
$K_{1,j}$ exhibits, as a function of $s$, a sharp cusp at $s=s_0$. An 
arbitrary variation of the input absorptive parts transforms an analytic input 
into a non-analytic one. We correct this partly and stay in the vicinity of an 
analytic input by modifying simultaneously the scattering length $a_2$ and 
choosing $\delta a_2$ in such a way that the cusp in $K_{1,j}$ is suppressed. 
This variation $\delta a_2$ is a linear functional of the $\delta A_i$:
\begin{equation}\label{4nine}
\delta a_2=\sum_{i=0}^2\int_{s_0}^\infty {\rm d}u\,\kappa_i(u)\delta A_i(u)
\end{equation}
and Section~3 tells us that equations of the form (\ref{4eight}) are still 
valid if the $K_{ij}$ are replaced by new kernels $\hat{K}_{ij}$.

The kernels $K_{ij}$ and $\hat{K}_{ij}$ have been evaluated as functions of 
$s$ for 3 values of $u$: $u_1=35$ ($E_1=828$~MeV) close to the matching point, 
$u_2=51$ ($E_2=1$~GeV) and $u_3=100$ ($E_3=1.4$~GeV).

\begin{center}
\begin{turn}{-90}
\epsfysize=8cm\epsfbox{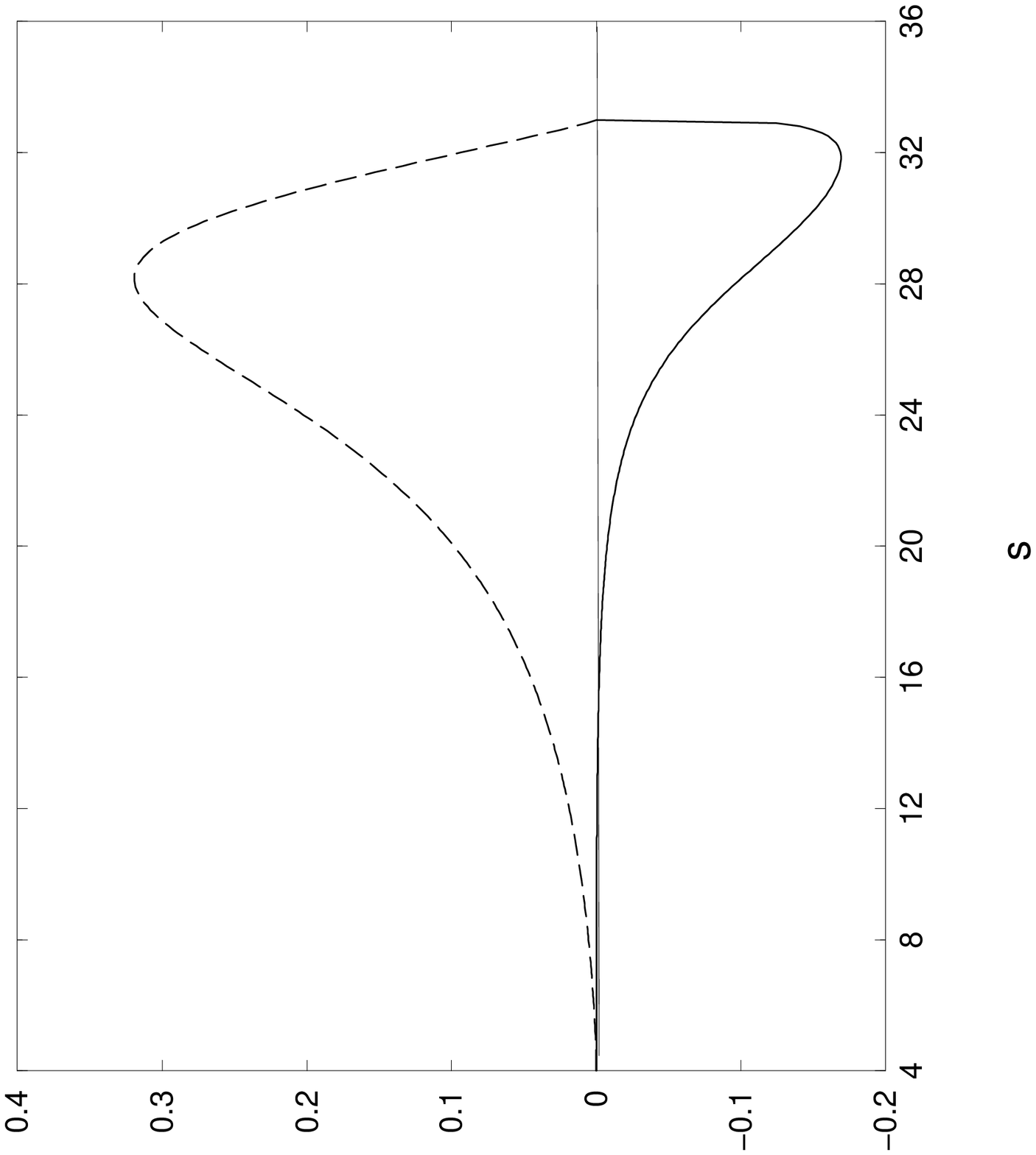}
\end{turn}
\end{center}
\setlength{\unitlength}{0.75mm}
{\small Figure 4: The kernels $K_{11}$ and $\hat{K}_{11}$ at $u=35$ as functions of $s$: 
$\hat{K}_{11}$ includes the effect of a variation of $a_0^2$ suppressing the 
cusp in $K_{11}$. 
\begin{picture}(14,2)(0,-2)\put(0,0){\line(1,0){14}}\end{picture} 
$K_{11},$ 
\begin{picture}(14,2)(0,-2)\multiput(0,0)(3,0){5}{\line(1,0){2}}\end{picture} 
$\hat{K}_{11}$.}

\vspace{4mm}
The passage from $K_{11}$ to $\hat{K}_{11}$ at $u=u_1=35$, slightly above the 
matching point, is illustrated in Fig.~4. As must be the case, the large cusp 
in $K_{11}$ has disappeared in $\hat{K}_{11}$. The effect of the induced 
variation of $a_2$ ($\kappa_1(u_1)=-0.0085$) dominates the response to the 
variation of $A_1$ outside the neighborhood of $s_0$. The fact that 
$\hat{K}_{11}$ is larger than $K_{11}$ is peculiar: $\hat{K}_{10}$ and 
$\hat{K}_{12}$ are much smaller than $K_{10}$ and $K_{12}$.

The values of the kernels $\hat{K}_{ij}$ at $u=u_1$ determine the responses to 
small variations $\delta A_j$ concentrated around that point. If $\delta A_j$ 
is sufficiently small and narrow the equations~(\ref{4eight}) and (\ref{4nine}) give
\begin{equation}\label{4ten}
\delta'_i(s)-\delta_i(s)\simeq\sum_j\hat{K}_{ij}(s,u_1)\Delta A_j,\qquad \delta 
a_2\simeq \sum_i\kappa_i(u_1)\Delta A_i,
\end{equation}
with
\begin{equation}\label{4elev}
\Delta A_i=\int{\rm d}u\,\delta A_i(u).
\end{equation}

The relative phase shift differences produced by such variations of the input 
absorptive parts with corresponding variations of the scattering 
length $a_2$ are displayed in Figs.~5, 6 and 7. The $\Delta A_i$ have been 
chosen in such a way that the responses are of the order of a few percent. Our 
linearization should be reliable under these circumstances. 
To describe the situation in physical terms we can imagine that the $\Delta 
A_j$ are produced by the insertion of fictitious narrow elastic resonances of 
width $\Gamma_j$ at $u_1$. The values of the $\Delta A_j$ used in Figs.~5, 6 
and 7 correspond to $\Gamma_0=0.76$~MeV, $\Gamma_1=1.56$~MeV, 
$\Gamma_2=0.92$~MeV. The effects of these resonances sitting just above the 
matching point spread over the whole interval $[4,s_0]$. The induced variation 
of $a_2$ produces a modest peaking of $(\delta'_2-\delta_2)/\delta_2$ at 
threshold. The responses to a variation $\Delta A_0$ in the isospin~0 S-wave 
are globally smaller than the effects of variations $\Delta A_1$ and $\Delta 
A_2$ of the same size in the other channels. A variation $\Delta A_i$ in channel $i$ produces a response in the same 
channel that is enhanced near $s_0$ and dominates the responses in the other 
channels. This dominance is significant but not very strong in the case of 
$\Delta A_2$. Apart from these observations we 
do not discover any striking feature characterizeing qualitatively the 
coupling of the S- and P-waves.

\begin{center}
\begin{turn}{-90}
\epsfysize=8cm\epsfbox{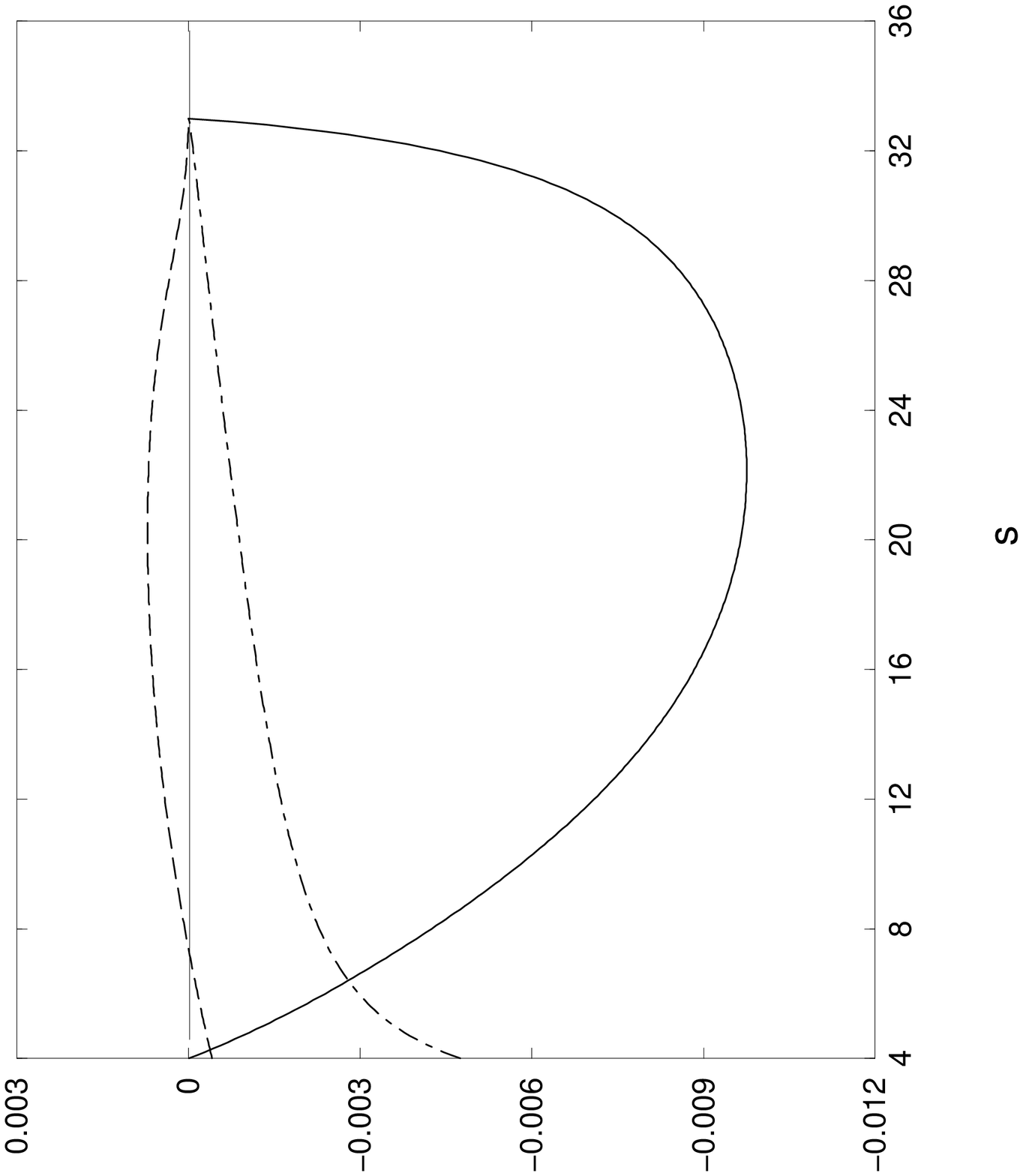}
\end{turn}
\end{center}
\nopagebreak
{\small Figure 5: Relative responses $(\delta_l^{I\,'}-\delta_l^I)/\delta_l^I$ to a 
variation of $A_0^0$ concentrated on $u=35$, $\Delta A_0^0=-0.1$.
\setlength{\unitlength}{0.75mm}
\begin{picture}(14,2)(0,-2)\put(0,0){\line(1,0){14}}\end{picture} 
$(\delta_0^{0'}-\delta_0^0)/\delta_0^0,$ 
\begin{picture}(14,2)(0,-2)\multiput(0,0)(3,0){5}{\line(1,0){2}}\end{picture} 
$(\delta_1^{1'}-\delta_1^1)/\delta_1^1,$  
\begin{picture}(14,2)(0,-2)\put(0,0){\line(1,0){2}}\put(3,0){\line(1,0){1}}
\put(5,0){\line(1,0){2}}\put(8,0){\line(1,0){1}}\put(10,0){\line(1,0){2}}
\put(13,0){\line(1,0){1}}
\end{picture} $(\delta_0^{2'}-\delta_0^2)/\delta_0^2$.}

\vspace{4mm}
\begin{center}
\begin{turn}{-90}
\epsfysize=8cm\epsfbox{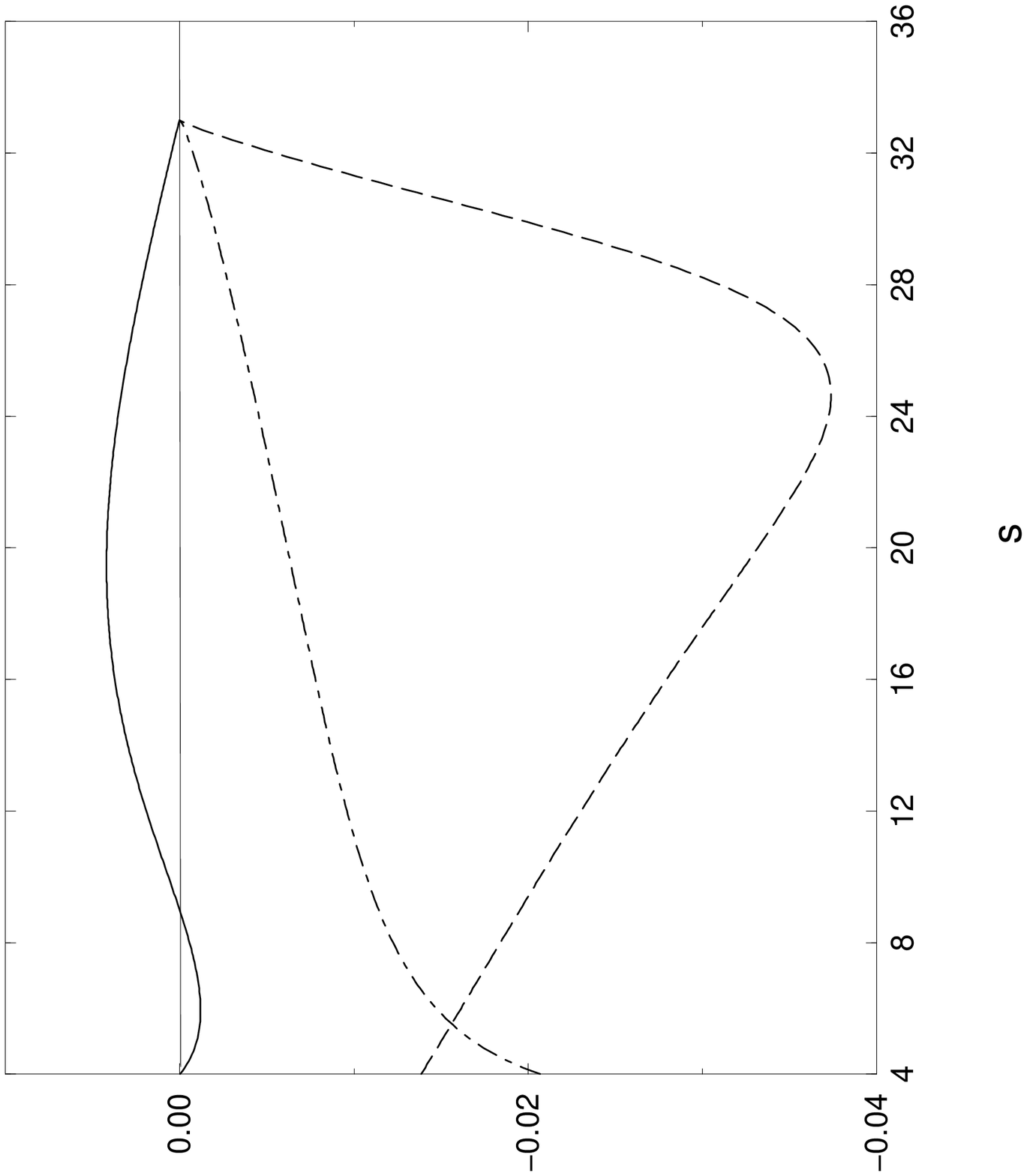}
\end{turn}
\end{center}
\nopagebreak
\setlength{\unitlength}{0.75mm}
{\small Figure 6: Relative responses $(\delta_l^{I\,'}-\delta_l^I)/\delta_l^I$ to a 
variation of $A_1^1$ concentrated on $u=35$, $\Delta A_1^1=-0.1$.
\begin{picture}(14,2)(0,-2)\put(0,0){\line(1,0){14}}\end{picture} 
$(\delta_0^{0'}-\delta_0^0)/\delta_0^0,$ 
\begin{picture}(14,2)(0,-2)\multiput(0,0)(3,0){5}{\line(1,0){2}}\end{picture} 
$(\delta_1^{1'}-\delta_1^1)/\delta_1^1,$  
\begin{picture}(14,2)(0,-2)\put(0,0){\line(1,0){2}}\put(3,0){\line(1,0){1}}
\put(5,0){\line(1,0){2}}\put(8,0){\line(1,0){1}}\put(10,0){\line(1,0){2}}
\put(13,0){\line(1,0){1}}
\end{picture} $(\delta_0^{2'}-\delta_0^2)/\delta_0^2$.}

\vspace{4mm}
\begin{center}
\begin{turn}{-90}
\epsfysize=8cm\epsfbox{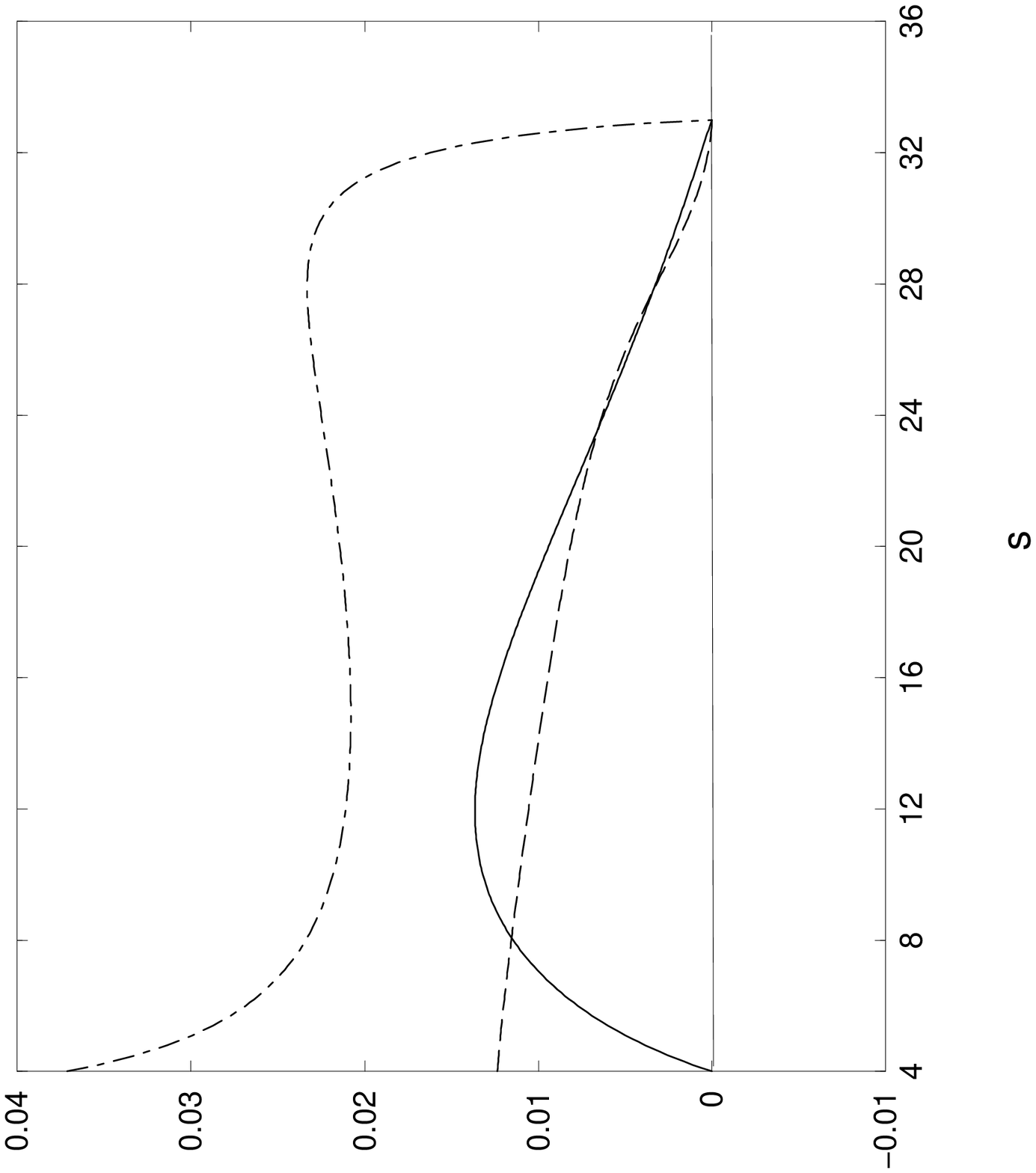}
\end{turn}
\end{center}
\setlength{\unitlength}{0.75mm}
{\small Figure 7: Relative responses $(\delta_l^{I\,'}-\delta_l^I)/\delta_l^I$ to a 
variation of $A_0^2$ concentrated on $u=35$, $\Delta A_0^2=0.1$. 
\begin{picture}(14,2)(0,-2)\put(0,0){\line(1,0){14}}\end{picture} 
$(\delta_0^{0'}-\delta_0^0)/\delta_0^0,$ 
\begin{picture}(14,2)(0,-2)\multiput(0,0)(3,0){5}{\line(1,0){2}}\end{picture} 
$(\delta_1^{1'}-\delta_1^1)/\delta_1^1,$ 
\begin{picture}(14,2)(0,-2)\put(0,0){\line(1,0){2}}\put(3,0){\line(1,0){1}}
\put(5,0){\line(1,0){2}}\put(8,0){\line(1,0){1}}\put(10,0){\line(1,0){2}}
\put(13,0){\line(1,0){1}}
\end{picture} $(\delta_0^{2'}-\delta_0^2)/\delta_0^2$.}

\vspace{4mm}
The $\hat{K}_{ij}$ are decreasing functions of $u$ without significant change 
in their shape as functions of $s$. The decrease is rapid just above the 
matching point. For instance, the $\hat{K}_{i0}$ are scaled down at $u=36$ to 
70\%\ of their values at $u=35$.

To characterize the decrease of the responses when variations $\Delta A_j$ are 
shifted to higher energies, we compute averages $\rho_{ij}$ of the absolute 
values of the relative phase shift differences at $u_1=35$, $u_2=51$ and 
$u_3=100$. According to (\ref{4ten}) these are given by 
\begin{equation}\label{4twelve}
\rho_{ij}(u_k)={1\over s_0-4}\int_4^{s_0}{\rm d}s\left|{\hat{K}_{ij}(s,u_k)\over 
\delta_i(s)}\Delta A_j\right|.
\end{equation}
Approximate values of the $\rho_{ij}(u_1)$ are given in Table~3. 
The ratios $\rho_{ij}(u_2)/\rho_{ij}(u_1)$ and 
$\rho_{ij}(u_3)/\rho_{ij}(u_1)$ show the decrease of the responses at higher 
energies. None of the mean responses to variations located at $u_2$ exceed 
$11\%$ of the corresponding responses at $u_1$. This percentage is reduced to 
$1.2\%$ when $u_2$ is replaced by $u_3$. Table~4 gives the values of 
the variations $\delta a_2$ coming from~(\ref{4ten}).

\begin{center}
\begin{tabular}{|c|c|ll|l|}
\hline
&&&&\\[-4mm]
$(i,j)$ & $\rho_{ij}(u_1)$ & \multicolumn{2}{|c|}{$\d {\rho_{ij}(u_2)\over 
\rho_{ij}(u_1)}$} & 
$\d {\rho_{ij}(u_3)\over \rho_{ij}(u_1)}$\\
&&&&\\[-4mm]
\hline
$(0,0)$ & $7.5\cdot 10^{-3}$ && 0.061 & 0.0049\\
$(1,0)$ & $4.2\cdot 10^{-4}$ && 0.10 & 0.010\\
$(2,0)$ & $1.1\cdot 10^{-3}$ && 0.10 & 0.010\\
\hline
$(0,1)$ & $2.4\cdot 10^{-3}$ && 0.10 & 0.012\\
$(1,1)$ & $2.6\cdot 10^{-2}$ && 0.073 & 0.0069\\
$(2,1)$ & $6.8\cdot 10^{-3}$ && 0.11 & 0.012\\
\hline
$(0,2)$ & $8.3\cdot 10^{-3}$ && 0.080 & 0.0069\\
$(1,2)$ & $7.5\cdot 10^{-3}$ && 0.079 & 0.0069\\
$(2,2)$ & $2.2\cdot 10^{-2}$ && 0.052 & 0.0035\\
\hline
\end{tabular}
\end{center}
{\small Table 3: Mean relative responses $\rho_{ij}(u_1)$ defined in (\ref{4twelve}) 
for $|\Delta A_j|=0.1$ and ratios of mean relative responses at $u_2$ and 
$u_3$ vs. responses at $u_1$, $\sqrt{u_1}\,M_\pi=828$~MeV, 
$\sqrt{u_2}\,M_\pi=1$~GeV, $\sqrt{u_3}\,M_\pi=1.4$~GeV.}
\vspace*{8mm}

We conclude that the solution of the Roy equations is quite insensitive to the 
errors on the input absorptive parts above $E_3=\sqrt{u_3}M_\pi=1.4$~GeV. 
The solution of the Roy 
equations is most sensitive to the input absorptive parts close to the 
matching point. According to Table~4, the uncertainty in $a_2$, associated in 
our scheme with an error on the absorptive parts at $u_3$, is less than $1\%$ 
of the uncertainty due to the same error at $u_1$.

\begin{center}
\begin{tabular}{|c|c|c|l|}
\hline
&&&\\[-4mm]
$j$ & $\d {\delta a_2(u_1)\over a_2}$ & $\d {\delta a_2(u_2)\over\delta 
a_2(u_1)}$ & $\d {\delta a_2(u_3)\over\delta 
a_2(u_1)}$\\
&&&\\[-4mm]
\hline
0 & $-4.8\cdot 10^{-3}$ & $0.09$ & $0.009$\\
1 & $-2.1\cdot 10^{-2}$ & $0.10$ & $0.011$\\
2 & $3.6\cdot 10^{-2}$ & $0.07$ & $0.006$\\
\hline
\end{tabular}
\end{center}
{\small Table 4: Relative variations of the scattering length $a_2$ induced according 
to (\ref{4ten}) by variations of the input absorptive parts $A_j$ at $u_1$, 
$|\Delta A_j|=0.1$ and ratios of variations at $u_2$ and $u_3$ vs. variations 
at $u_1$, $\sqrt{u_1}\,M_\pi=828$~MeV, $\sqrt{u_2}\,M_\pi=1$~GeV, 
$\sqrt{u_3}\,M_\pi=1.4$~GeV.}
\vspace*{8mm}

We close the discussion of variations of the input absorptive parts with an 
assessment of the accuracy of our results. The 
errors come from our functions $H_i$. These form an approximate solution of 
equations~(\ref{Bone}) with inhomogeneous terms $Z_i$ containing a component 
(\ref{4three}) coming from variations of the $A_j$ at $u_k$ and a component 
(\ref{3six}) due to the corresponding variation of $a_2$. Let $\chi_j(u_k)$ be 
the total discrepancy between left- and right-hand sides of eq.~(\ref{Bone}) 
defined in (\ref{Belev}). These quantities are listed in Table~5. 
All equations (\ref{Bone}) are verified at least at the percent level, which is 
sufficient for our purpose.

\begin{center}
\begin{tabular}{|l|l|l|l|}
\cline{2-4}
\multicolumn{1}{l|}{ } &
\multicolumn{1}{c|}{$u_1$} &
\multicolumn{1}{c|}{$u_2$} &
\multicolumn{1}{c|}{$u_3$}\\
\hline
$\chi_0$ & 0.012 & 0.012 & 0.005 \\[2mm]
$\chi_1$ & 0.024 & 0.057 & 0.016 \\[2mm]
$\chi_2$ & 0.017 & 0.019 & 0.014 \\
\hline
\end{tabular}
\end{center}
{\small Table 5: Total discrepancies $\chi_j$ defined in (\ref{Belev}) to variations 
of the input absorptive part $A_j$ at $u_k$ and the correlated variation of 
the scattering length $a_2$, $\sqrt{u_1}\,M_\pi=828$~MeV, 
$\sqrt{u_2}\,M_\pi=1$~GeV, $\sqrt{u_3}\,M_\pi=1.4$~GeV.}
\vspace*{8mm}

We close this section with a survey of the response to variations of the 
driving terms $\psi_i$ in equation~(\ref{2two}). The $\psi_i$ are small and 
approximated by polynomials on $[4,s_0]$ in~\cite{anant}. We consider 
variations of these polynomials. The Ansatz for the functions $h_i$ defined in 
(\ref{2eight}) becomes
\begin{equation}\label{4thirt}
h_i(s)=(s-4)G_i(s)\left(p\,\delta_{i,1}+H_i(s)\right),
\end{equation}
where $p$ is a constant and the $H_i$ form a solution of the 
equations~(\ref{Bone}) with
\begin{equation}\label{4fourt}
Z_i(s)=\delta\psi_i(s)+\delta_{i,1}\,p\int_4^{s_0}{\rm 
d}x\,R_{i1}(s,x)\,\sin(2\delta_1(x))\,G_1(x).
\end{equation} 
As before, the variations of the driving terms are combined with variations of 
$a_2$ such that $h_1$ has no cusp at $s_0$. The result shows that large 
relative variations of the driving terms affect only weakly the phase shifts 
below $s_0$. For instance, a reduction of the size of $\psi_0$ or $\psi_2$ by 
50$\%$ changes the $\delta_i$ by less than $5\%$. In the case of a $50\%$ 
reduction of $\psi_1$ the response is smaller than $0.5\%$.  

\section{Summary and conclusions}
We have developed an approximation scheme to determine the linear response of 
the solution of the S- and P-wave Roy equations with matching point $s_0=33$ 
to small variations of their input (S-wave scattering lengths, S- and P-wave 
absorptive parts above $s_0$, and driving terms). Our results are precise at 
the percent level, which is sufficient for a qualitative insight. Our problem 
has been solved long ago, in a different way, for a higher matching point 
$s_0=70$ in~\cite{epelew}. At $s_0=33$ the solution of the Roy equations is 
unique, entirely determined by their input.

An arbitrary input leads to a solution that is singular at $s_0$. As the 
physical amplitudes are regular at $s_0$, the physical input belongs to the 
restricted class of our analytic inputs producing a solution that is 
non-singular at $s_0$. We prove that under legitimate assumptions an analytic 
input has in fact only one solution regular at $s_0$ (Appendix~A).

An arbitrary variation of the input transforms an analytic input into a 
non-analytic one and induces responses that are singular at $s_0$. Due to the 
fact that our $s_0$ is close to $M_\rho^2$ ($M_\rho=\rho$-meson mass), the 
sharpest singularities show up as cusps in the isospin~1 P-wave responses. 
These cusps are suppressed by correlating suitably the variations of two 
pieces of the input. We choose to associate in this way variations of the 
isospin~2 S-wave scattering length $a_0^2$ to arbitrary variations of other 
components of the input.

It is instructive to compare our strategy with the procedure used 
in~\cite{anant} when solving the Roy equations themselves. The solution is 
parametrized in~\cite{anant} by an Ansatz that is regular at $s_0$. As one is 
working with a non-analytic approximation of the physical input, the solution 
is singular at $s_0$ and cannot be fitted exactly by the Ansatz. An 
approximate solution is constructed by a least square procedure tuning 
simultaneously the parameters in the Ansatz and the scattering length $a_0^2$ 
in the input. In this way the input is brought close to an analytic one and 
the Ansatz gives a model of the corresponding solution. In some of its 
features this machinery resembles our simple strategy. In fact, their 
equivalence for the computation of responses to small variations of the input 
has been checked in the case of the variation of the isospin~0 S-wave 
absorptive parts displayed in Fig.~5. The response obtained by solving the 
full Roy equations coincides with our result within a few percents. This confirms that 
the main factor tuning $a_0^2$ in~\cite{anant} is the avoidance of a cusp in 
the isospin~1 P-wave phase shift.

Our technique shows that one stays in the vicinity of an analytic input when 
moving infinitesimally along a given direction of the $(a_0^0,a_0^2)$ plane 
without changing the other pieces of the input. This confirms the existence of 
a so-called universal curve at the linear response level.

We have determined the response to localized variations of the input 
absorptive parts above the matching point. It spreads over the whole interval 
$[4,s_0]$ and illustrates the intricate coupling of the S- and P-waves 
produced by crossing symmetry. It shows that the sensitivity to the errors in 
the input absorptive parts decreases rapidly with increasing energy.

{\bf Acknowledgments} I thank G.~Colangelo, J.~Gasser and H.~Leutwyler for 
their continual interest in the present work. I am especially grateful to 
J.~Gasser for his help in the preparation of the figures.

\appendix
\section{Analytic input and uniqueness}

An analytic input $(a_i,A_i,\psi_i,\eta_i)$ is defined as an input admitting 
at least one solution of the Roy equations which is regular at the matching 
point. A precise definition is given below. In any case it is an indirect 
definition: as shown at the end of this appendix, we know how to construct 
analytic inputs but we are unable to identify an analytic input by direct 
inspection. Its components are correlated: in 
particular, the scattering lengths depend on the $A_i$, $\psi_i$ and $\eta_i$. 
Analytic inputs are relevant objects because the physical input belongs to 
that class. The aim of this appendix is to prove that an analytic input has 
only one solution regular at $s_0$. The requirement of regularity at $s_0$ 
eliminates in principle the uniqueness problem. This result has already been 
established in~\cite{gasserw} for simplified one-channel Roy equations.

To establish our result we need general analyticity properties of the partial 
wave amplitudes $f_i$. Let $f$ be one of them. It is known to be the boundary 
value of an analytic function $F$ on the interval $[4,125.31]$~\cite{royw}. 
This function is holomorphic in a complex domain $\Delta$ extending on the real 
axis from $s_{\rm L}=-28$ to $s_{\rm R}=125.31$ and provided with a left-hand 
cut $[s_{\rm L},0]$ and a right-hand cut $[4,s_{\rm R}]$. We have
\begin{equation}\label{Aone}
f(s)=\lim_{\epsilon\searrow 0}F(s+{\rm i}\epsilon),\qquad s\in[4,s_{\rm R}].
\end{equation}

Our matching point $s_0$ being above the first inelastic threshold $i_1=16$, 
we need properties characterizing the elasticity parameters $\eta$ which 
enter into an analytic input. According to (\ref{2five}), $\eta$ is equal to 
the modulus of the S-matrix element $(1+2{\rm i}\sigma f)$ which is the 
boundary value of
\begin{equation}\label{Atwo}
S(z)=1-2\,\sqrt{4-z\over z}F(z).
\end{equation}
This function is regular in $\Delta$. Using the relation 
$\bar{S}(z)=S(\bar{z})$ we write
\begin{equation}\label{Athree}
\eta^2(s)=\lim_{\epsilon\searrow 0}S(s+{\rm i}\epsilon)\bar{S}(s+{\rm 
i}\epsilon)=\lim_{\epsilon\searrow 0}S(s+{\rm i}\epsilon)S(s-{\rm 
i}\epsilon),\; s\in [i_1,s_{\rm R}].
\end{equation}
Although it cannot be derived from first principles~\cite{martin}, it is legitimate to 
assume that the inelastic thresholds $i_k$ ($k=1,2,\dots$) are the only 
singularities of $f$ on $[4,s_{\rm R}]$ and that $S$ has an analytic 
continuation $S_{II}$ into the sheet reached by crossing the cut $[4,s_{\rm 
R}]$ from below between two successive inelastic thresholds $i_k$ and 
$i_{k+1}$ [$S_{II}$ depends on the pair $(i_k,i_{k+1})$]. 
Equation~(\ref{Athree}) gives 
\begin{equation}\label{Afour}
\eta^2(s)=\lim_{\epsilon\searrow 0}S(s+{\rm i}\epsilon)S_{II}(s+{\rm 
i}\epsilon)
\qquad s\in (i_k,i_{k+1}),
\end{equation}
as long as $i_{k+1}<s_{\rm R}$.

We assume that $S_{II}$ is regular in the upper half-plane, in a neighborhood 
$D$ of the segment $(i_k,i_{k+1})$. Equation~(\ref{Afour}) tells 
us that the real-valued $\eta^2$ is the boundary value on $(i_k,i_{k+1})$ of a 
function holomorphic in $D$ and we apply the following general result.

\newtheorem{lem}{Lemma}
\begin{lem} Let $w$ be a real-valued function defined on the interval 
$(i_k,i_{k+1})$. If $w$ is the boundary value of an analytic funtion $W$ 
holomorphic in $D$, it is the restriction to $(i_k, i_{k+1})$ of a function 
regular in the domain $D\cup\bar{D}$ where $\bar{D}$ is the mirror domain of 
$D$: $\bar{D}=\{z|\bar{z}\in D\}$.
\end{lem}

A proof of this Lemma is given at the end of this appendix. It implies that 
$\eta^2$ has an analytic continuation regular in a 
complex neighborhood of $(i_k,i_{k+1})$. We assume that the possible complex 
zeros of $\eta^2$ are at a finite distance from $(i_k,i_{k+1})$. We choose $D$ 
sufficiently narrow so that $\eta^2$ is non-vanishing on $D$ and we have

\begin{lem} If the above conditions are fulfilled $\eta$ has a holomorphic 
continuation from each interval $(i_k,i_{k+1})$ with $i_{k+1}<s_{\rm R}$ into 
a complex neighborhood of that interval with $i_k$ and $i_{k+1}$ on its 
boundary.
\end{lem}

We turn now to properties of the full amplitude $f$ and establish

\begin{lem} The real and imaginary parts of $f$ are separately holomorphic in 
a complex neighborhood of each interval $(i_k,i_{k+1})$ ($i_{k+1}<s_{\rm R}$) 
with $i_k$ and $i_{k+1}$ on its boundary. Here $k=0,1,2,\dots$ with $i_0=4$.
\end{lem}

This is a well known result in the case of the interval 
$[4,i_1]$~\cite{zimmermann}. For any interval we define the function
\begin{equation}\label{Afive}
V={1\over {\rm i}\sigma}\,{1-\eta+2{\rm i}\sigma f\over 1+\eta+2{\rm i}\sigma f}
\end{equation}
on $(i_k,i_{k+1})$ [$\eta=1$ on $(4,i_1)$]. According to Lemma~2, $V$ has a 
regular analytic continuation into a domain $N$ in the upper half plane --
~$[i_k,i_{k+1}]$ belongs to the boundary of $N$ --~except for poles at the 
possible zeros of the denominator. Using unitarity
\begin{equation}\label{Asix}
\Im\,f=\sigma|f|^2+{1\over 4\sigma}(1-\eta^2)
\end{equation}
we find that $\Im\,V=0$ on $(i_k,i_{k+1})$. Lemma~1 is easily extended to the 
case of meromorphic functions and one concludes that $V$ has a meromorphic 
continuation into $N\cup\bar{N}$. The definition~(\ref{Afive}) gives
\begin{equation}\label{Aseven}
\Re\,f={\eta V\over 1+\sigma^2V^2},\qquad \Im\,f={\sigma\eta V^2\over 
1+\sigma^2V^2}+{1\over 2\sigma}(1-\eta).
\end{equation}
We assume again that the zeros of the denominators are at a finite distance 
from the real axis and discover that $\Re\,f$ and $\Im\,f$ are indeed 
separately holomorphic in a neighborhood of $(i_k,i_{k+1})$ contained in 
$N\cup\bar{N}$. The phase shift 
$\delta$ is also regular in a neighborhood of each $(i_k,i_{k+1})$.

We close our preliminaries with the structure of $f$ at an inelastic threshold 
$i_k$ with square root singularity. There are four 
functions $a$, $b$, $c$ and $d$ that are regular in a circle $C_k$ with center 
$i_k$ and radius $\rho$ such that
\begin{equation}\label{Aeight}
\begin{array}{l}
\eta_>(s)=\exp\left[-2\left(a(s)+\sqrt{s-i_k}\,b(s)\right)\right]\\
\delta_>(s)=c(s)+\sqrt{s-i_k}\,d(s)
\end{array}
\end{equation}
for $s\in(i_k,i_{k+1})$~: $\eta_>$ and $\delta_>$ designate respectively the 
elasticity parameter and the phase shift above $i_k$. The amplitude $f$ can 
be written on $(i_k,i_{k+1})$ in terms of a complex phase shift 
$\tilde{\delta}_>$ as 
\begin{equation}\label{Anine}
\begin{array}{rcl}
\d f&=&{1\over 2{\rm i}\sigma}
\left({\rm e}^{2{\rm i}\tilde{\delta}_>}-1\right),\\
\tilde{\delta}_>(s)&=&c(s)+{\rm i}a(s)+\sqrt{s-i_k}\,(d(s)+{\rm i}b(s)).
\end{array}
\end{equation}

The value of $f$ below $i_k$, on $(i_k-\rho,i_k)$, is obtained through 
analytic continuation of the expression~(\ref{Anine}) in the upper half-plane 
along curves contained in $C_k$. The outcome is determined by a complex phase 
shift
\begin{equation}\label{Aten}
\tilde{\delta}_<(s)=c(s)+{\rm i}a(s)-\sqrt{i_k-s}\,(b(s)-{\rm i}d(s)).
\end{equation}
The elasticity parameter $\eta_<$ and the phase shift $\delta_<$ below $i_k$ 
are given by
\begin{equation}\label{Aelev}
\begin{array}{l}
\d \eta_<(s)=\exp\left[-2\left(a(s)+\sqrt{i_k-s}\,d(s)\right)\right],\\
\d \delta_<(s)=c(s)-\sqrt{i_k-s}\,b(s),
\end{array}\end{equation}
on $(i_k-\rho,i_k)$. The functions $b$ and $d$ interchange their roles when we 
cross $i_k$. Below $i_1$, $\eta_<$ is equal to 1. This implies
\begin{equation}\label{Atwel}
a=d=0
\end{equation}
in the case $k=1$.

We summarize our findings in
\begin{lem} The structure of $f$ at a square root inelastic threshold $i_k$ is 
described by formulas (\ref{Aeight}) and (\ref{Aelev}). Equation (\ref{Atwel}) 
holds at $i_1$.
\end{lem}

After this lengthy preparation we are ready for a complete definition of an 
analytic input.

{\bf Definition 1} An analytic input $(a_i,A_i,\psi_i,\eta_i)$ contains 
elasticities fulfilling Lemma~2 on $(4,s_0)$. It admits at least one solution 
$f_i$, $i=0,1,2$, of the S- and P-wave Roy equations with $\Re f_i$ regular 
at $s_0$ in the sense that they are holomorphic in a circle $C_{s_0}: |s-
s_0|<\epsilon$. These $f_i$ satisfy Lemmas~3 and 4.

We establish the following

\newtheorem{prop}{Proposition}
\begin{prop} \label{prop1}
Let $f_i$, $i=0,1,2$, form a solution of the S- and P-wave Roy 
equations with analytic input $(a_i,A_i,\psi_i,\eta_i)$ that is regular at 
$s_0$ and verifies Lemmas~3 and 4. A second solution $f_i'$ of these 
equations, $f'_i\neq f_i$, is singular at $s_0$.
\end{prop}

This proposition is an extension of Proposition~4 in~\cite{gasserw} to the 
realistic situation. In the following proof we assume $i_1<s_0<i_2$, which is 
true for our $s_0=33$.

The proof of Proposition~\ref{prop1} is based on Lemmas~2, 3 and 4. To make 
sure that the Roy equations~(\ref{2one}) guarantee the required analyticity of 
the $f_i$ we rewrite these equations as follows
\begin{equation}\label{Athirt}
\Re\,\Phi_i(s)=(s-4){1\over \pi}\Pint_4^\infty{\rm d}x\,{1\over (x-4)(x-
s)}\Im\,\Phi_i(x),
\end{equation}
where
{\small 
\begin{equation}\label{Afourt}
\Phi_i(s)=f_i(s)-a_i-(s-4)\left\{c_i(2a_0-5a_2)+\sum_{j=0}^2{1\over \pi}\int_4^\infty{\rm 
d}x\,R_{ij}(s,x)\Im\,f_j(x)+\psi_i(s)\right\}.
\end{equation}}

In these equations $\Im\,f_i(s)=A_i(s)$ for $s\geq s_0$, and the $\psi_i$ are 
the driving terms appearing in (\ref{2two}). The fact that 
$\Im\,\Phi_i=\Im\,f_i$ on $[4,\infty)$ has been used. The equations 
(\ref{Athirt}) ensure that 
the $\Phi_i$ are boundary values of analytic functions holomorphic in 
$\C\backslash[4,\infty)$. For $x\in[4,\infty)$ the kernels $R_{ij}(s,x)$ are 
holomorphic functions of $s$ in $\C\backslash(-\infty,0]$ and the driving 
terms are regular in the domain $\Delta$ without right-hand 
cut~\cite{royw}. 
Taking all this into account, (\ref{Afourt}) provides a representation of the 
$f_i$ ensuring that they are indeed boundary values of functions, 
holomorphic in the domain $\Delta$, with right- and left-hand cuts. The same 
conclusion holds for the $f'_i$.

To establish Proposition~1, we show that the $f'_i$ have to coincide with the 
$f_i$ if the  $\Re\,f'_i$ are regular at $s_0$. Inversion of the dispersion 
relations (\ref{Athirt}) gives
\begin{equation}\label{Afift}
\Im\,\Phi_i(s)=\Im\,f_i(s)=-(s-4){1\over \pi}\Pint_\R {{\rm d}x\over x-
4}{\Re\,\Phi_i(x)\over x-s},\;s\in[4,\infty).
\end{equation}

The regularity of $\Re\,f_i$ at $s_0$ implies that 
$\Re\,\Phi_i$ is regular at that point and it follows from (\ref{Afift}) that 
$\Im\,f_i$ is holomorphic in $C_{s_0}$. The relation (\ref{Afift}) holds true 
for $\Phi'_i$ defined by $f'_i$ and the assumed regularity of $\Re\,f'_i$ at 
$s_0$ implies the analyticity of $\Im\,f'_i$ in $C_{s_0}$. According to 
Definition~1, $\Im\,f_i$ and $\Im\,f'_i$ have analytic continuations 
holomorphic in a complex neighborhood $N'$ of the interval $[s_,s_0]$ shown 
in Fig.~8. We conclude that $\Im\,f_i$ and $\Im\,f'_i$ are holomorphic in 
$N'\cup C_{s_0}$. As
\begin{equation}\label{Asixt}
\Im\,f'_i(s)=\Im\,f_i(s)=A_i(s) \mbox{ for }s\in[s_0,s_0+\epsilon),
\end{equation}
$\Im\,f'_i$ and $\Im\,f$ coincide in $N'\cup C_{s_0}$ and
\begin{equation}\label{Aseventbis}
\Im\,f_i(s)=\Im\,f'_i(s) \mbox{ for }s\in(i_1,s_0].
\end{equation}

\begin{center}
\setlength{\unitlength}{0.4mm}
\begin{picture}(236,80)(-110,-40)
\put(0,0){\circle{40}}
\put(96,0){\circle{40}}
\put(-52,0){\ellipse{104}{80}} 
\put(48,0){\ellipse{96}{72}} 
\put(-104,0){\circle*{2}}
\put(0,0){\circle*{2}}
\put(96,0){\circle*{2}}
\put(-116,0){\line(1,0){236}}
\put(-110,-8){4}
\put(2,-8){$i_1$}
\put(98,-6){$s_0$}
\put(46,16){$N'$}
\put(-50,20){$N''$}
\put(-12,5){$C_1$}
\put(98,5){$C_{s_0}$}
\end{picture}
\end{center}

{\small Figure 8: Domains of the complex $s$-plane used in the proof of 
Proposition~1.}

\vspace{4mm}
To complete our proof, we have to extend the equality (\ref{Asixt}) below the 
inelastic threshold $i_1$. The discussion of Lemma~4 shows that phase shifts 
are required to go through $i_1$. The equality of imaginary parts above $i_1$ at 
fixed $\eta_i$ implies
\begin{equation}\label{Asevent}
\delta'_i(s)=\pm\delta_i(s)\quad\mod\pi,\qquad s\in(i_1,s_0].
\end{equation}
We show that the minus sign must be rejected. If, for a given $i$, 
$\delta'_i=-\delta_i\;\mod\pi$, (\ref{Aeight}) gives
\begin{equation}\label{Aninetbis}
\delta'_{i>}(s)=c'(s)+\sqrt{s-i_1}\,d'(s)
\end{equation}
with $c'(s)=-c(s)\;\mod\pi$, $d'(s)=-d(s)$. According to (\ref{Aelev}) this 
would produce an elasticity $\eta'_{i<}$ below $i_1$ that would differ from 
the input elasticity $\eta_{i<}$.

In terms of the functions $b$ and $c$ appearing at $i_1$, we now have
\begin{equation}\label{Aninet}
c'_i=c_i\quad\mod\pi
\end{equation}
whereas $b'_i=b_i$ is given, $\eta_i$ being a member of the input. 
Equations~(\ref{Aten}) and (\ref{Atwel}) give
\begin{equation}\label{Atwen}
\delta'_{i<}(s)=\left(c_i(s)-\sqrt{i_k-s}\,b_i(s)\right)\;\mod\pi,\quad s\in(i_1-
\rho,i_1).
\end{equation}
Therefore we have
\begin{equation}\label{Atwone}
\Im\,f'_i(s)=\Im\,f_i(s),\qquad s\in(i_1-\rho,i_1).
\end{equation}
As both sides of this equations have analytic extensions regular in a 
neighborhood $N''$ of $(4,i_1)$ the equality~(\ref{Atwone}) extends to 
$(4,i_1)$. 
Thus we have established that $f'_i$ and $f_i$ have the same imaginary parts on 
$[4,s_0]$ if $i_1<s_0<i_2$ and the Roy equations imply the full equality of 
these two amplitudes. This result extends to arbitrary choices of the matching 
point. The proof becomes easier if $s_0<i_1$; it requires more steps if 
$s_0>i_2$.

For completeness we prove Lemma~1. We define a function $\hat{W}$ by 
$\hat{W}(z)=\bar{W}(\bar{z})$. It is holomorphic in the mirror domain 
$\bar{D}$ and, $w$ being real, we have $\d w(s)=\lim_{\epsilon\searrow 
0}\,\hat{W}(s-{\rm i}\epsilon)$, $s\in[i_k,i_{k+1}]$. We write for $z\in D$
\begin{equation}\label{Atwthree}
W(z)={1\over 2{\rm i}\pi}\orintg{D}{\rm d}x\,{W(x)\over x-z}+{1\over 
2{\rm i}\pi}\orintg{\bar{D}}{\rm d}x\,{\hat{W}(x)\over x-z}.
\end{equation}
The first term is the Cauchy representation of $W$ and the second integral 
vanishes because $z\notin\bar{D}$. The contributions of the segment 
$[i_k,i_{k+1}]$ to both integrals cancel and one is left with the Cauchy 
representation of a function holomorphic in $D\cup\bar{D}$.

Three remarks close this appendix.
\begin{enumerate}
\item If $\sqrt{s_0}M_\pi=800$~MeV, we know, according to Section~2, that the 
physical solution of the Roy equations is an isolated one. The relevance of 
Proposition~1 comes from the possible existence of other solutions with 
$\delta'_i(s_0)=\delta_i(s_0)+n_i\pi$ resulting from CDD-pole 
ambiguities~\cite{atkinson}. These solutions are singular at $s_0$.
\item The proof of Proposition~1 tells us that the absorptive parts $A_i$ of 
an analytic input are regular on some interval $[s_0,s'_0)$ above the matching 
point ($s'_0\geq s_0+\epsilon$) and are the analytic continuation of 
$\Im\,f_i$ below $s_0$ on that interval. The Roy equations (\ref{2one}) define 
real parts $\Re\,f_i$ above $s_0$. On $[s_0, s'_0)$ they are the analytic 
continuations of the $\Re\,f_i$ below $s_0$. As the interval $[s_0,s'_0)$ 
cannot contain an inelastic threshold, all the ingredients of the unitarity 
condition (\ref{Asix}) have analytic continuations from below $s_0$ onto 
$[s_0,s'_0)$. This implies that (\ref{Asix}) holds on $[s_0,s'_0)$: $\Re\,f_i$ 
and $A_i$ are the real and imaginary parts of amplitudes verifying unitarity 
on that interval. This means that they fulfill, at least on $[s_0,s'_0)$, a 
consistency condition discussed in~\cite{anant}.
\item Although we have no direct way of checking whether a given input is an 
analytic one, we have a recipe for the construction of such inputs. Take a 
matching point $s'_0$ above $s_0$ ($s'_0<125.31$) and choose arbitrarily an 
input $(a'_i,A'_i,\psi'_i,\eta'_i)$. Let $f'_i$ be a solution of the Roy 
equations with that input, verifying Lemmas~2, 3 and 4. These $f'_i$ are 
expected to be singular at $s'_0$ but they are regular at $s_0$. Define a new 
Ansatz $(a_i,A_i,\psi_i,\eta_i)$ with matching point $s_0$:
\begin{eqnarray}
a_i&=&a'_i, \nonumber\\
\psi_i&=&\psi'_i, \quad \eta_i=\eta'_i \quad\mbox{ on }[4,s_0],\nonumber\\
A_i(s)&=& \left\{\begin{array}{ll} 
\Im\,f'_i(s) &\mbox{for } s_0\leq s\leq s'_0,\\
A'_i(s) &\mbox{for } s>s'_0. \end{array}\right. \label{Atwfour}
\end{eqnarray}
The $f'_i$ define a solution $f_i$ of this new problem,
\begin{equation}\label{Atwfive}
f_i(s)=f'_i(s)\mbox{ for }4\leq s\leq s_0.
\end{equation}
This solution is regular at $s_0$ and the Ansatz $(a_i,A_i,\psi_i,\eta_i)$ is 
an analytic one.

Our recipe is of no practical use because it 
requires the explicit resolution of the Roy equations with matching 
point $s'_0$. The important point is that we recognize that an analytic input 
with matching point $s_0$ is unconstrained above some $s'_0$, $s'_0>s_0$. It 
is the behavior of the $A_i$ on $[s_0,s'_0]$ which is constrained and $s'_0$ 
can be close to $s_0$.

In our definition an input is analytic with respect to its matching point 
$s_0$. The physical input is special because it generates inputs with matching 
points $s'_0>s_0$ ($s'_0<125.31$) that are analytic with respect to $s'_0$.
\end{enumerate}

\setcounter{equation}{0}
\section{Approximation scheme}
We write the equations we have to solve in Sections~3 and 4 in the 
following way.
\begin{equation}\label{Bone}
\sum_{j=0}^2 X_{ij}[H_j](s)=Z_i(s).
\end{equation}
Each $X_{ij}$ is a linear and homogeneous functional of the unknown $H_j$,
\begin{eqnarray}
X_{ij}[H_j](s)\!\!&\!\!=\!\!&\!\!\delta_{i,j}\left\{\delta_{m_{j,0}}H_j(s)
-{1\over\pi}\int_4^{s_0}{\rm d}x\,G_j(x)\sin(2\delta_j(x)){H_j(x)-H_j(s)\over 
x-s}\right\}\nonumber \\
&&\qquad -{1\over\pi}\int_4^{s_0}{\rm d}x\,R_{ij}(s,x)G_j(x)\sin(2\delta_j(x))H_j(x).
\label{Btwo}\end{eqnarray}
The $Z_i$ are known functions determined by the variation of the input under 
consideration. The unknown $H_i$ are regular and slowly varying on $[4,s_0]$ 
and we approximate them by polynomials
\begin{equation}\label{Bthree}
H_i(s)=s^{m_i}\sum_{n=0}^Nc_{i,n}\,s^n.
\end{equation}
We have to determine the coefficients $c_{i,n}$. The $X_{ij}$ become
\begin{equation}\label{Bfour}
X_{ij}[H_j](s)=\sum_{n=0}^NX_{ij}^{(n)}(s)c_{j,n}
\end{equation}
where the $X_{ij}^{(n)}$ are known functions obtained by replacing $H_j(x)$ by 
$x^{(m_j+n)}$ in the right-hand side of Eq.~(\ref{Btwo}).

To evaluate these functions we define auxiliary analytic functions 
$\bar{G}_i$, holomorphic in $\C\backslash[4,s_0]$
\begin{equation}\label{Bfive}
\bar{G}_i(z)=\left({s_0\over s_0-z}\right)^{m_i}\exp\left[{2\over\pi}
\int_4^{s_0}{\rm d}x{\delta_i(x)\over x-z}\right].
\end{equation}
They are related to the $G_i$ defined in (\ref{2eleven}) by their 
discontinuity Disc~$\bar{G}_i$ across the cut $[4,s_0]$,
\begin{equation}\label{Bsix}
{1\over 2{\rm i}}{\rm Disc}\,\bar{G}_i(s)=G_i(s)\sin(2\delta_i(s)),\quad 4\leq 
s\leq s_0.
\end{equation}
The contribution to $X_{ij}^{(n)}$ coming from the first integral in the 
right-hand side of (\ref{Btwo}) is transformed into a sum of integrals along a 
closed contour $\Gamma$ surrounding the segment $[4,s_0]$:
\begin{equation}\label{Bseven}
-{1\over 2{\rm i}\pi}\sum_{m=0}^{n-1}\orinth{\rm 
d}z\,\bar{G}_i(z)z^ms^{n-m-1}=-\sum_{m=0}^{n-1}g_{i,m+1}s^{n-m-1}
\end{equation}
where the $g_{i,p}$ are the coefficients of the Laurent series of $\bar{G}_i$,
\begin{equation}\label{Beight}
\bar{G}_i(z)=\sum_{p=0}^\infty g_{i,p}{1\over z^p}.
\end{equation}
The second integral in the right-hand side of (\ref{Btwo}) is evaluated in a 
similar way by exploiting the analyticity properties of the kernels $R_{ij}$. 
At fixed real $s$, $s\geq 4$, these are analytic functions of $x$, holomorphic 
in $\C\backslash[-(s-4),0]$. Deforming the contour $\Gamma$, we get
\begin{eqnarray}
\lefteqn{\int_4^{s_0}{\rm d}x\,R_{ij}(s,x)\bar{G}_j(x)\sin(2\delta_j(x))x^n 
=}\qquad &&\nonumber\\
&&-{1\over 2{\rm i}\pi}\int_{-(s-4)}^0{\rm d}x\,{\rm 
Disc}\,R_{ij}(s,x)\,\bar{G}_j(x)x^n +\mbox{ polynomial.}\label{Bnine}
\end{eqnarray}
The polynomial is determined by the asymptotic behavior in $z$ of the product 
$R_{ij}(s,z)G_j(z)z^n$. The discontinuity Disc~$R_{ij}$ of $R_{ij}$ across 
$[-(s-4),0]$ being known, we need the $\bar{G}_j$ on that interval. These 
smooth 
functions are approximated by third degree polynomials at a level smaller than 
1\%. This allows the explicit evaluation of the integral in the right-hand 
side of Eq.~(\ref{Bnine}) and the result is a polynomial in $s$. The 
$X_{ij}^{(n)}$ are thus approximated by polynomial $\tilde{X}_{ij}^{(n)}$ of 
degree $\leq 6$.

Evaluated along the same lines, the inhomogeneous terms $Z_i$ in 
Eqs.~(\ref{Bone}) become known functions $\tilde{Z}_i$. According to the 
equations~(\ref{Bone}) the $\tilde{X}_i$ and $\tilde{Z}_i$ have to be made 
approximately equal on $[4,s_0]$ by adjusting the $3(N+1)$ coefficients 
$c_{n,i}$ in Eq.~(\ref{Bfour}) [$\tilde{X}_i$ is obtained by substituting 
$\tilde{X}_{ij}^{(n)}$ for $X_{ij}^{(n)}$ in (\ref{Bfour}) and inserting the 
result into (\ref{Bone})]. We keep our calculations simple by using polynomials 
of low degree for the $H_i$ in (\ref{Bthree}) and choose $N=2$. To determine 
the 9 coefficients $c_{i,n}$, the $\tilde{X}_i$ and $\tilde{Z}_i$ are 
approximated on $[4,s_0]$ by second degree polynomials using a $\chi^2$ 
technique, and these polynomials are set equal. This gives 9 equations for the 
9 unknowns (in $\tilde{X}_i$, each $\tilde{X}_{ij}^{(n)}$ is replaced by a 
polynomial of degree 2). The whole procedure is legitimate because the 
$\tilde{X}_i$ and $\tilde{Z}_i$ are slowly varying.

The $\tilde{X}_{ij}^{(n)}$ and $\tilde{Z}_i$ are close to the $X_{ij}^{(n)}$ 
and $Z_i$, the differences coming only from the replacement of the $\bar{G}_i$ 
by third degree polynomials on $[-(s-4),0]$. Thus, in view of 
equations~(\ref{Bone}), the $c_{i,n}$ we obtain must be such that 
$\tilde{X}_i$ and $\tilde{Z}_i$ are close to each other on $[4,s_0]$. This can 
be checked by evaluating the mean relative quadratic discrepancies of 
$\tilde{X}_i$ and $\tilde{Z}_i$ 
\begin{equation}\label{Bten}
\chi_i^{\mbox{ }}=\left[\d {1\over (s_0-4)}\mbox{\large$\d \int_4^{s_0}$}{\rm d}s{\left(\tilde{X}_i(s)-\tilde{Z}_i(s)\right)^2
\over \left(\tilde{Z}_i(s)\right)^2}\right]^{1\over 2}.
\end{equation}
We can also define a total discrepancy
\begin{equation}\label{Belev}
\chi=\left[{1\over 3}\sum_{i=0}^2\chi_i^2\right]^{1\over 2}.
\end{equation}
The various values we obtain for these quantities are quoted in Sections~3 
and 4.

\setcounter{equation}{0}
\section{The kernels $R_{ij}$}
Our technique makes extensive use of the analyticity properties of the regular 
kernels $R_{ij}$ in equations~(\ref{2one}) and (\ref{2two}). It is therefore 
convenient to display them explicitly. They are obtained from 4 functions 
$L_k$, $k=1,\dots,4$:
{\small
\begin{eqnarray}
L_1(s,x)\!\!&\!\!=\!\!&\!\!{1\over x(s-4)}\left[{1\over 2}s-x+2+(x-4){x\over s-4}
\ln\left(1+{s-4\over x}\right)\right],\nonumber\\[4mm]
L_2(s,x)\!\!&\!\!=\!\!&\!\!{1\over x(s-4)}\left[-{3\over 2}s-x+2+(2s+x-4){x\over s-
4}\ln\left(1+{s-4\over x}\right)\right],\nonumber\\[4mm]
L_3(s,x)\!\!&\!\!=\!\!&\!\!{1\over x(s-4)^2}\left\{-{1\over 6}\left[s^2-8s+4(3x^2-
12x+4)\right]\right.\label{Cone}\\
\!\!&\!\!\!\!&\!\!\hspace*{2.5cm}\left.+(2s+x-4)(x-4){x\over s-
4}\ln\left(1+{s-4\over x}\right)\right\},\nonumber\\[4mm]
L_4(s,x)\!\!&\!\!=\!\!&\!\!{1\over x(s-4)^2}\left\{-{1\over 6}\left[s^2+8s(3x-1)+4(3x^2-
12x+4)\right]\right.\nonumber\\
\!\!&\!\!\!\!&\!\!\hspace*{4cm}\left.+(2x+s-4)(2s+x-4)(x-4){x\over s-
4}\ln\left(1+{s-4\over x}\right)\right\}.\nonumber
\end{eqnarray}
}

The $R_{ij}$ are given by
\begin{equation}\label{Ctwo}\begin{array}{ll}
\d R_{00}(s,x)={2\over 3}L_1(s,x)-{1\over x}, &
\d R_{02}(s,x)={10\over 3}L_1(s,x),\\[2mm]
\d R_{20}(s,x)={2\over 3}L_1(s,x), &
\d R_{22}(s,x)={1\over 3}L_1(s,x)-{1\over x},\\[2mm]
\d R_{01}(s,x)=6L_2(s,x), &
\d R_{21}(s,x)=-3L_2(s,x),\\[2mm]
\d R_{10}(s,x)={2\over 3}L_3(s,x), &
\d R_{12}(s,x)=-{5\over 3}L_3(s,x),\\[2mm]
\d R_{11}(s,x)=3L_4(s,x)-{1\over x}.
\end{array}
\end{equation}


\end{document}